\documentclass[graybox, nosecnum]{svmult}

\usepackage{mathptmx}       
\usepackage{helvet}         
\usepackage{courier}        
\usepackage{type1cm}        
\usepackage{makeidx}         
\usepackage{graphicx}        
\usepackage{multicol}        
\usepackage[bottom]{footmisc}
\usepackage{hyperref}        
\usepackage{soul}            

\makeindex             

\def\lQ{\Lambda_{\rm QCD}}
\def\als{\alpha_{\rm s}} 
\def\bfnabla{\mbox{\boldmath $\nabla$}}
\def\bfsigma{\mbox{\boldmath $\sigma$}}
\def\lQ{\Lambda_{\rm QCD}}
\def\als{\alpha_{\rm s}} 
\def\siml{{\ \lower-1.2pt\vbox{\hbox{\rlap{$<$}\lower6pt\vbox{\hbox{$\sim$}}}}\ }}

\newcommand{\be}{\begin{equation}}
\newcommand{\ee}{\end{equation}}
\newcommand{\bea}{\begin{eqnarray}}
\newcommand{\eea}{\end{eqnarray}}
\newcommand{\nn}{\nonumber}

\begin{document}
\title*{Quark Nuclear Physics  with Heavy Quarks}
\author{Nora Brambilla \thanks{corresponding author} }
\institute{Nora Brambilla \at    Physik-Department, Technische Universit\"at M\"unchen,
  James-Franck-Str. 1, 85748 Garching, Germany;
  Institute for Advanced Study,     Technische Universit\"at M\"unchen, Lichtenbergstrasse 2~a, 85748 Garching, Germany;
   Munich Data Science Institute, Technische Universit\"at M\"unchen,  
Walther-von-Dyck-Strasse 10, 85748 Garching, Germany
  \email{nora.brambilla@ph.tum.de}}
%
%
\maketitle  
\abstract{Heavy quarks have been instrumental for progress in our exploration of strong interactions.
 Quarkonium in particular, a  heavy quark-antiquark nonrelativistic  bound state, has been at the root
 of   several revolutions.   Quarkonium  is  endowed with  a pattern of separated energy scales 
qualifying it as special probe of complex environments.   Its multiscale nature  has  made  a  description  in
Quantum  Field  Theory particularly  difficult  up  to the  advent  of nonrelativistic effective field theories.
We will focus on systems made by two or more heavy quarks. After considering some historical 
approaches based on the potential models and the Wilson loop approach, we will introduce the contemporary 
nonrelativistic effective field theory descriptions, in particular potential nonrelativistic QCD which 
entails the Sch\"odinger equation as zero  order problem, define the potentials as matching coefficients
and allows systematic calculations of the physical properties. 
The effective field theory  allows us to explore quarkonium properties in the realm of QCD. In particular it allows us
to make calculations with unprecedented precision
when high order perturbative calculations are possible and to systematically factorize  short from long range contributions where
observables are sensitive to the nonperturbative  dynamics of QCD.
Such  effective field theory treatment  can be extended at  finite temperature and in presence of gluonic and light quark excitations.
We  will show that  in this novel theoretical framework,
quarkonium  can play a crucial role for a number of problems at the frontier of our research, from the investigation of 
the confinement dynamics in strong interactions to the study of deconfinement and the phase diagram of nuclear matter,
to the precise determination of Standard Model parameters up to the 
emergence of exotics X Y Z states of an unprecedented nature.}

\section{\textit{The role of heavy quarks}}

Heavy quarks and bound states of heavy quarks, primarily quarkonium, a bound state of a heavy quark and a heavy antiquark,
have been historically instrumental to construct the theory of strong interactions and  continue today to be at the forefront
of our research as a golden  probe  of the strong dynamics.

The discovery of heavy quarks drastically changed the Standard Model  (SM) of particle physics. This happened in 
the November revolution of 1974 when  two labs  on opposite sides of USA announced 
discovery of a new particle, a fact  that helped the  acceptance of the Standard Model of particle physics
\cite{E598:1974sol,SLAC-SP-017:1974ind}.
The new particle was the first example of quarkonium: the  $J/\psi$,  the lowest excitation made by a charm and an anticharm.
The $J/\psi$  appeared as an unprecedented sharp peak, tall and narrow,  3 GeV in mass and
90  KeV in width, at variance with  the typical width of several tens and hundreds of  MeV of the hadrons discovered
up to that time, i.e. strongly interacting
light quark  composite particles.
   The $J/\psi$ discovery represented the confirmation of the quark model,
   the discovery of the charm quark, the confirmation of the GIM mechanism \cite{Glashow:1970gm}
   (the mechanism through which flavour-changing neutral currents are suppressed in loop diagrams) and
   the first discovery of a quark of large mass moving nonrelativistically.
   It triggered additional searches and   in  few years the  higher excitations of
   charmonium were discovered  as well has bottomonia (1977), i.e.
   bound states of bottom and antibottom,   $B_c$  (1998),  and the  top (1995).
   States made by top and antitop decay weakly before forming a proper bound state, however still leaving their
   signature in the form of an enhancement of the cross section at threshold \cite{QuarkoniumWorkingGroup:2004kpm}.

   The  $J/\psi$ discovery was  the confirmation of QuantumChromoDynamics (QCD) \cite{Fritzsch:1973pi}, the Quantum Field Theory
   describing strong interactions. QCD has
    a well defined behaviour in the ultraviolet (UV) region at large energy
   and a fundamental coupling constant $\alpha_s=g^2/4 \pi$ running  from small values at large energy to large values at small energy.
   This encodes the properties of asymptotic freedom (quarks are free at high momentum transfer)
   and confinement (quarks are confined in color singlet hadrons
   at low energy) \cite{Wilson:1974sk}. As we will see,  confinement becomes manifest in the case of heavy quarks, where one can write 
   the color singlet quark-antiquark interaction potential in terms of a so called Wilson loop   \cite{Brambilla:1999ja}.
   Confinement emerges in an area law of the Wilson loop 
   and correspondingly in a linear potential growing with the distance  between the quarks \cite{Creutz:1976ch}.
    An emergent scale  $\Lambda_{QCD}$ parametrizes the importance of the nonperturbative corrections, i.e. the contributions
   that cannot be calculated in an expansion in $\alpha_s$ and is mirrored in the hadron spectrum, the mass of the
   proton  being  proportional to    $\Lambda_{QCD}$. We consider a quark to be heavy when its mass $m$ is larger than the scale
   $\Lambda_{QCD}$: this qualifies as heavy the charm, bottom and top quarks.

       These fundamental features of QCD  find the best  realization   in the $J/\psi$ and in  quarkonium in general.
    The small width can be explained by the fact that $J/\psi$ is the lowest energy level and can decay only via annihilation,
    which makes available in the process a large energy, of order of two times the mass of the charm (about 2 GeV). The annihilation
    width    is then proportional to $\alpha_s^2(2 m_c)$ which is small due to asymptotic freedom, since $m_c$ is bigger than  $\Lambda_{QCD}$.
    On the other hand, when theorists set up to investigate the structure of  the
    energy levels of  charmonium and bottomonium, 
     they noticed that it can be reproduced by using in
     the Schr\"odinger equation     a      static potential superposition of an attractive Coulomb contribution (with the
     appropriate $SU(3)$
     color factor for a singlet $Q\bar{Q}$)
     and a term linear in the distance: the famous Cornell potential   \cite{Eichten:1978tg,Eichten:1979ms}.
     Such form of the potential has been later confirmed by 
     nonperturbative calculations performed using computational lattice  QCD
     \cite{Creutz:1976ch,Brambilla:1999ja,Rothe:1992nt,Bali:2000gf,Campostrini:1987ht}.  In 1977  the static potential 
     was calculated at two loops \cite{Fischler:1977yf} which enabled to reobtain the QCD $\beta$ function at order $O(g^5)$ and the solution
     of the Callan-Symanzik equation to order $g^5$,  which gave a hint on the existence of a color  confining potential.

In the following we will address the importance of heavy quarks for  quark nuclear physics.
We will focus on systems made by two or more heavy quarks and  discuss how they play a crucial
role for a number of problems at the frontier of our research, from the investigation of 
the confinement dynamics in QCD to the study of deconfinement and the phase diagram of nuclear matter,
to the precise determination of Standard Model parameters up to the 
emergence of exotics X Y Z states of an unprecedented nature
\cite{Brambilla:2010cs,QuarkoniumWorkingGroup:2004kpm,Andronic:2015wma,Chapon:2020heu,Brambilla:2014jmp}.
In particular, we will conclude that our  progress in these strong interactions topics is connected to a
broad sweep of physical problems  in settings ranging from astrophysics and cosmology 
to strongly   correlated systems in particle and condensed matter physics as well as to search of physics beyond the Standard model.

Heavy quarks  had and have a key role  to address  
the phenomenon of CP (Charge Conjugation and  Parity) violation, which is crucial to understand the asymmetry between matter and antimatter
that exists in the Universe,  by testing and constraining 
the Cabibbo-Kobayashi-Maskawa matrix entries entering 
$B$ and $D$ mesons decays and mixing and   by identifying new physics contributions, 
see e.g. the reviews \cite{Buchalla:2008jp,Neubert:1996qg,Chang:2017wpl,Gershon:2016fda}.
 The Babar and Belle experiments at the B factories \cite{Belle-II:2018jsg,BaBar:2014omp} have been constructed to this aim 
 but turned out to be also formidable heavy mesons machines giving a great boost to out knowledge of heavy quark systems
 and their strong interaction dynamics.

 The review is organized as follows. 
 First, we  will  discuss the physics characteristics of systems made by heavy quarks. Then,  we will summarize what
 have been historically the pioneering  approaches and the phenomenological models. 
 After that, we will explain that to address in quantum field theory a nonrelativistic multiscale system like quarkonium,
 it is necessary resort to nonrelativistic effective field theories (NREFTs).
We will show how to construct NREFTs up to arriving at the simplest possible version,
called potential nonrelativistic QCD (pNRQCD)  which  implements the Schr\"odinger equation as the zero order problem.
Then, we will show how  combining  NREFTs and lattice we can get systematic and under control predictions on
a number of physical processes and observables like the spectrum, decays, transitions and production.
In this framework quarkonium becomes  a  unique laboratory for the study of strong interactions from the
high energy to the low energy scales.
The NREFT allows us to explore quarkonium properties in the realm of QCD. In particular it allows us
to make calculations with unprecedented precision
when high order perturbative calculations are possible and to systematically factorize  short from range contributions where
observables are sensitive to the nonperturbative  dynamics of QCD.
 Finally, we will address some of the most interesting open problems in relation to quarkonium, i.e. the non-equilibrium 
 propagation of quarkonium in medium which calls for introducing open quantum system on top of NREFTs and
 the new exotic stares X Y Z discovered at the accelerator experiments.

\section{\textit{Heavy-light mesons, quarkonia , baryons with two or more heavy quarks  }}

\begin{figure}[ht]
\centering
\includegraphics[width=0.9\linewidth]{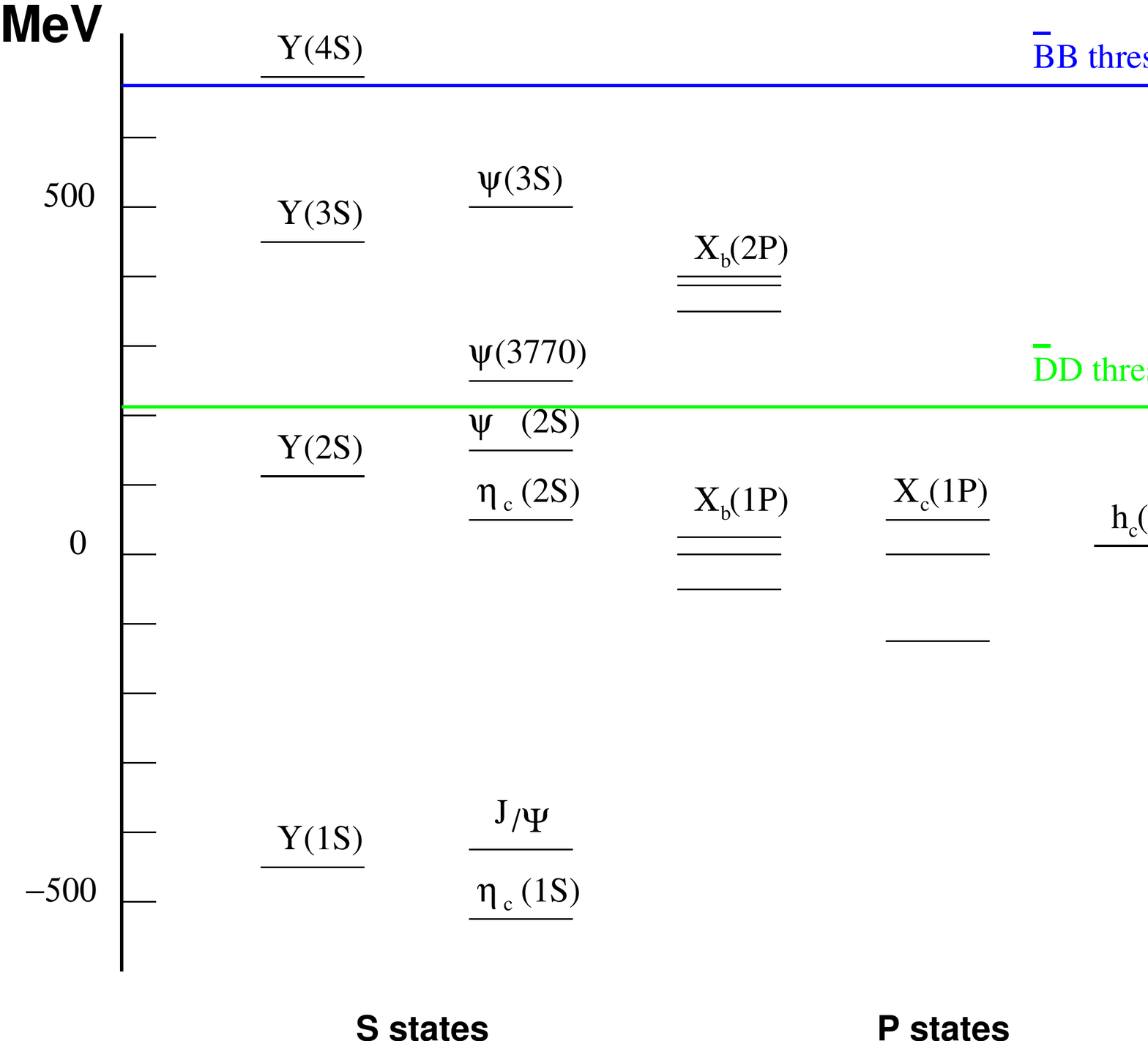}
\caption{The experimental quarkonium energy levels ($b\bar{b}$ and $c\bar{c}$) as known in the eighties (plotted as relative to the 
  spin-average of the $\chi_b(1P)$ and $\chi_c(1P) $ states to be able to present the two sectors in the same plot).
  The states  are identified by names and mass in
  parenthesis, $\Upsilon$ being the vector states (with $L=0$) in bottomonium, $J/\psi$
  and $\psi$  the vector states in charmonium sector, $\eta$ the pseudoscalar states and $\chi$ and $h$ 
  states with $L=1$ and different total angular momentum. The first strong decay thresholds are also shown.}
\label{spect}
\end{figure}

\begin{figure}[ht]
\centering
\includegraphics[width=0.9\linewidth]{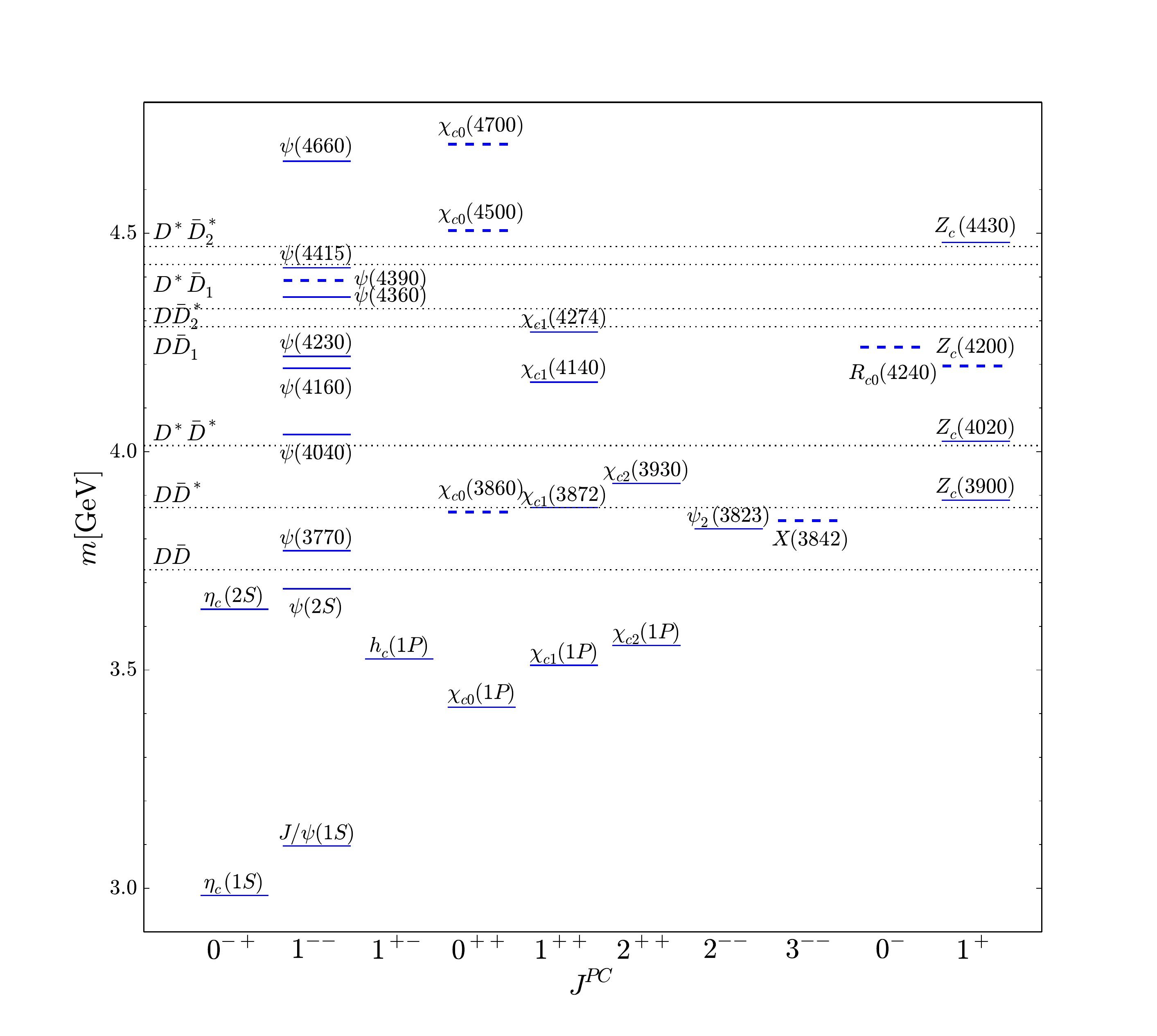}
\caption{The spectrum of states in the $c\bar{c}$ sector as of 2019, taken from ref. \cite{Brambilla:2019esw}. 
Thin solid lines represent the states established experimentally and dashed lines are for those that are claimed but not (yet) established
(a state is regarded as established if it is seen in different modes).
States whose quantum numbers are undetermined are not shown.
States in the plot are labeled according to the PDG primary naming scheme that superseded the X Y Z notation, for further details  see \cite{Brambilla:2019esw}. 
Dashed lines show some relevant thresholds that open in the considered mass range; here $D_1$ stands for $D_1(2420)$ and $D_2^*$ for $D_2^*(2460)$.
Thresholds with hidden strangeness or involving broad states are not shown.
The states shown in the two columns to the right are isovectors containing a $\bar c c$ pair; they are necessarily exotic. 
The just discovered state $X(6900)$ \cite{LHCb:2020bwg}   made by two charm and two anticharm quarks  would be out of the scale of the figure. More states have been 
observed after 2019.
} 
\label{cc}
\end{figure}

\begin{figure}[ht]
\centering
\includegraphics[width=0.9\linewidth]{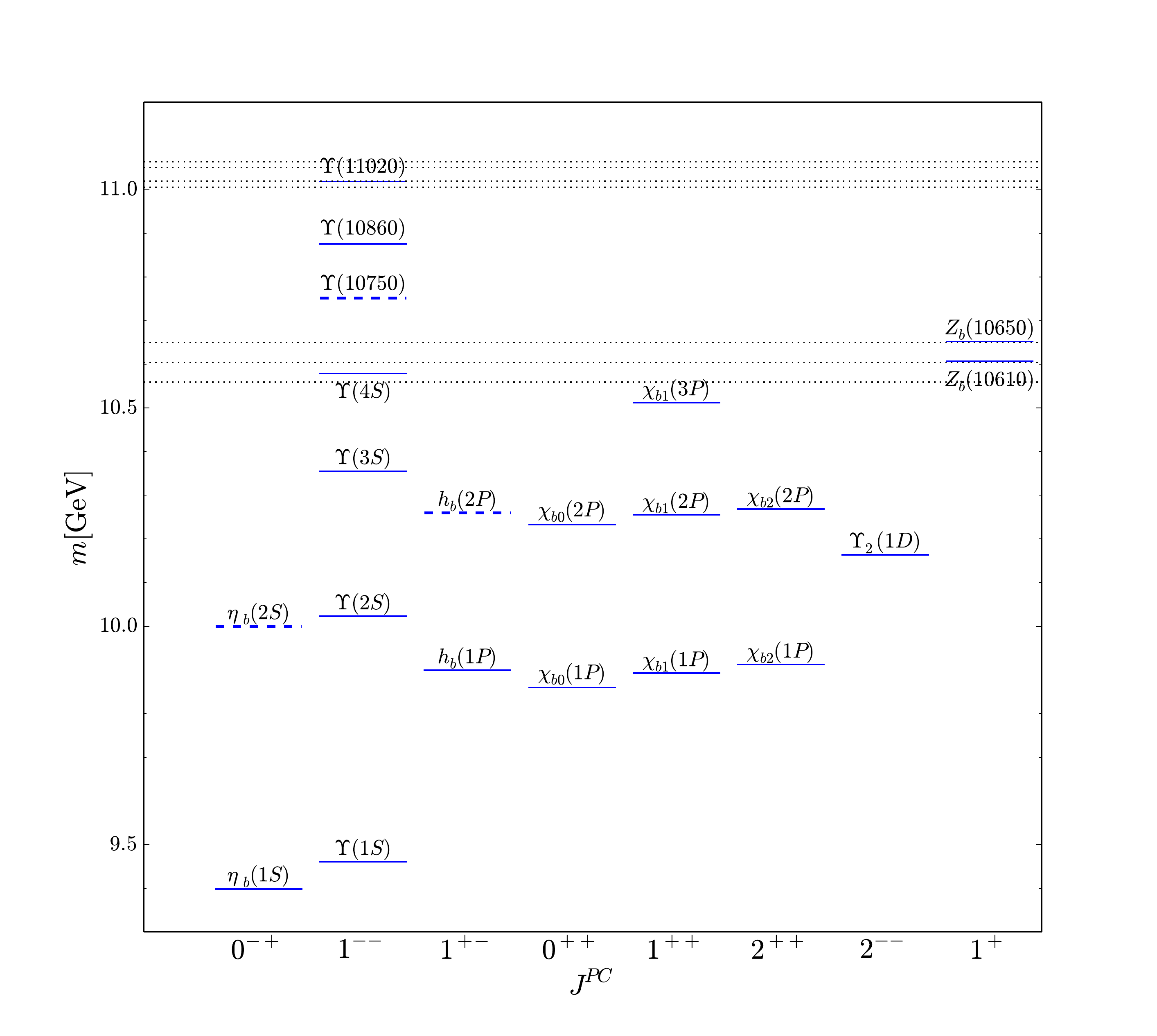}
\caption{The spectrum of states in the $\bar cc$ sector as of 2019, taken from ref. \cite{Brambilla:2019esw}. 
Thin solid lines represent the states established experimentally and dashed lines are for those that are claimed but not (yet) established
(a state  is regarded as established if it is seen in different modes).
States whose quantum numbers are undetermined are not shown.
States in the plot are labeled according to the PDG primary naming scheme that superseded the X Y Z notation, for further details  see \cite{Brambilla:2019esw}. 
Dashed lines show some relevant thresholds that open in the considered mass range; here $D_1$ stands for $D_1(2420)$ and $D_2^*$ for $D_2^*(2460)$.
Thresholds with hidden strangeness or involving broad states are not shown.
The states shown in the two columns to the right are isovectors containing a $\bar c c$ pair; they are necessarily exotic. }
\label{bb}
\end{figure}

Heavy quarkonia  are systems composed by a heavy quark and a heavy antiquark
with mass $m$ larger than the ``QCD  confinement scale'' $\lQ$, 
so that $\als(m) \ll 1$ holds. We have that $m_c \sim 1.5$ GeV and $m_b \sim 5$ GeV.
Of course the quark masses are scheme dependent objects and their actual value depends
on the considered scheme
and scale,  only the pole mass being scale independent, see e.g. the review \cite{Brambilla:2004jw}.
From the quarkonia spectra, see Fig. \ref{spect}  (and \ref{cc} and \ref {bb}),
it is evident that 
the difference in the orbital energy levels is much smaller than the quark mass.
It scales like  $mv^2$, while  fine and hyperfine separations scale like $mv^4$. Here
$v$ is the heavy quark velocity
( $v = |\vec{v}|$)  in the rest frame of the meson {in units of $c$}, and 
$v^2 \sim 0.1$ for the $b\bar{b}$, $v^2 \sim 0.3$ for $c\bar{c}$ systems. This is the same
scaling of the hydrogen atom if one identifies $v$ with the fine structure constant $\alpha_{\rm em}$.
Therefore quarkonia are nonrelativistic 
systems. 
 Being nonrelativistic, quarkonia 
are characterized  by a hierarchy of energy scales: the mass $m$ of the heavy quark (hard scale),
the  typical relative momentum $p \sim m v$ (in the meson rest frame) corresponding to the 
inverse Bohr radius $r \sim 1/(m v)$ (soft scale),
 and the typical binding  energy $E \sim m v^2$ (ultrasoft scale). 
Of course, for quarkonium there is another scale that can never be switched off in QCD, i.e.
$\lQ$, the scale at which non perturbative effects become dominant.
A similar pattern of scales emerge in the case of 
baryons composed of  two or three heavy quarks \cite{Brambilla:2005yk} and  for the
just discovered state $X(6900)$ \cite{LHCb:2020bwg}   made by two charm and two anticharm quarks.
The pattern  of nonrelativistic scales makes all the 
difference between heavy quarkonia
and heavy-light mesons, which are characterized
by just two scales: $m$ and $\lQ$.

Being a multiscale system, heavy quarkonium is  probing different energy regimes of 
the strong interactions, from the hard region, where an expansion in the 
coupling constant is possible and precision studies may be done, to the 
low-energy region,  dominated by confinement and the many manifestations  of 
nonperturbative dynamics. In addition, the properties of production  
and absorption  of quarkonium in  a nuclear  and hot  medium  are crucial inputs for 
the study of QCD at high density  and temperature, reaching out to cosmology.
On the experimental side the diversity, quantity and accuracy of the data  collected in the last decades
  is impressive  and includes clean and precise samples of quarkonia spectra decay, transition and  production
processes, including the  discovery of exotics  X Y Z states  at  tau-charm  (BES experiment)
and B factories (Babar and Belle experiments),
hadroproduction at Fermilab Tevatron and the Large Hadron Collider (LHC) experiments at CERN, production in photon-gluon 
fusion at DESY, photoproduction at Jlab,
heavy ions production and suppression at RHIC, NA60, and  LHC. New data are coming also from
the upgraded experiments
BELLE II and BESIII \cite{Brambilla:2010cs,QuarkoniumWorkingGroup:2004kpm,Andronic:2015wma,Chapon:2020heu,Yuan:2019zfo}
and more will come in future from  Panda at FAIR and the Electron Ion Collider (EIC) \cite{Yuan:2019zfo,PANDA:2021ozp}.
On the theoretical side  the last few decades   have seen 
the construction of new nonrelativistic effective field 
theories and new  developmentd in computational lattice QCD which supply 
us with a systematic calculational framework in quantum field theory.
All this make quarkonium a  golden probe of strong interactions.

From now on we will concentrate 
on the study of systems with two or more heavy quarks.
We will outline the rich interplay of theoretical advancement and 
experimental success and its implication
on our control of strong interactions   inside the Standard Model of Particle Physics.  
 Before, however, I will summarize  the models that have been used in the past
 to describe the quarkonium properties.

\subsection{\textit{ The potential and the phenomenology of quarkonium}}

You find in Figs. (\ref{cc}) and (\ref{bb})  our present knowledge about the states in the charmonium and in the bottomonium sector.
We will consider in the following quarks of equal mass for simplicity (in the case of $B_c$ one should take into account that
the masses are different).
The quark and the antiquark spins combine to give the total spin 
${\bf S}={\bf S}_1 + {\bf S}_2$ which combines with the orbital angular momentum ${ \bf L}$ to give 
the total angular momentum ${ \bf J}$. The resulting state solution of a Schr\"odinger equation
is denoted by $n ^{2 S+1}L_J$ where 
$n-1$ is the number of radial nodes of the wave function. As usual, to $L=0$ is given the name $S$, to $L=1$ the name $P$, 
to $L=2$ the name D and so on. The experimental resonances (see  Figs. (\ref{cc}) and (\ref{bb}))  are  classified via the $J^{PC}$ quantum numbers, 
$P= (-1)^{L+1} $ being  the parity number and $C= (-1)^{L+S}$  the C-parity.
Strong decay thresholds are marked by horizontal dashed lines
that represent the energy necessary to decay in a  couple of heavy-light mesons.
In the charmonium sector only  10 states of the states presented in Fig. (\ref{cc})
have been discovered before  1980 and no one between 1980 and 2002. In 2003  the
new revolution started with the discovery of the $X(3872)$  \cite{Belle:2003nnu}
and a number of new states above and below the strong decay threshold
\cite{Brambilla:2010cs}.  Many of the states discovered at or above threshold
presented exotic features and have been initially termed X Y Z states
 \cite{Brambilla:2019esw}.

 The first tool used to describe quarkonium in the eighties has been the quark model with some notions of
 QCD incorporated. In particular, constituent heavy quark masses have been considered and a static potential
 called Cornell potential \cite{Eichten:1978tg,Eichten:1979ms}  has been used in a Schr\"odinger equation,
 successfully reproducing the quarkonia levels measured at that time. 
The Cornell potential  is flavor independent and has the form
\begin{equation}
V_0(r)= -  {\kappa  \over r}+ \sigma r + {\rm const},
\label{cornell}
\end{equation}
$r$ being the modulus of the quark-antiquark distance. 
The parameters 
$\kappa$   should be identified with $ {4\over 3} \alpha_{\rm s}$, corresponding to the one gluon exchange that
  should dominate at small distances  due to asymptotic freedom, and the string tension $\sigma$  corresponds
  to a constant energy density related to confinement and originating a potential growing with the interquark distance at large distances.
A fit to the states gave  $\kappa = 0.52$ and $\sigma=0.182$ GeV$^2$.
Since then, several different phenomenological forms of the static potential have been exploited,
see \cite{Lucha:1991vn,Brambilla:1999ja} for a review.
In order to explain the fine and hyperfine structure of the quarkonium spectrum, however, spin dependent relativistic corrections
 to the static potential have to  be considered. Moreover, for $v^2 \sim 0.1$ for the $b\bar{b}$, $v^2 \sim 0.3$ for $c\bar{c}$ systems,
one expects relativistic corrections of order $20\div 30 \%$ for the charmonium spectrum 
and up to $10 \%$ for the bottomonium spectrum and also spin independent but momentum dependent  corrections  have to be considered,
arriving at a phenomenological hamiltonian of the type
\begin{equation}
H= \sum_{j=1,2} \left(m_j + {p_j^2 \over 2 m_j} -{p_j^4 \over 8 m_j^3}\right) + V_0 + V_{\rm SD} + V_{\rm VD}.
\label{ham}
\end{equation}
The $1/m^2$  spin-dependent $V_{\rm SD} $ and velocity-dependent $V_{\rm VD} $ potentials 
were initially derived in the eighties  from  the semirelativistic reduction of a  Bethe--Salpeter (BS)
\cite{Salpeter:1951sz}   equation for the quark-antiquark 
connected amputated Green functions  or, equivalently at this level, from the  semirelativistic 
reduction of the quark-antiquark scattering amplitude with an effective exchange equal to the BS kernel.  
Several ambiguities are involved in this procedure, due on  one hand  to the fact that 
we do not know the relevant confining nonperturbative Bethe--Salpeter kernel, on the other hand  due to the fact 
that we have to get rid of the temporal (or energy $Q_0$, $Q=p_1-p_1^\prime$ being the momentum transfer) 
dependence of the kernel to recover a potential (instantaneous) description.
It turned out that, 
at the level of the approximation involved, the spin-independent relativistic corrections at the order $1/ m^2$ 
depend  on the way in which $Q_0$ is fixed together with
the gauge choice of the kernel \cite{Brambilla:1992fx,Lucha:1991vn,Brambilla:1999ja}.
The Lorentz structure of the kernel was also not known.  
On a phenomenological basis, the following ansatz for the kernel was intensively studied \cite{Brambilla:1999ja}
\begin{eqnarray}
I(Q^2)= (2 \pi)^3 \left[ \gamma_1^\mu  \gamma_2^\nu P_{\mu\nu}  J_v(Q) + J_s(Q) \right]
\label{kernel}
\end{eqnarray}
in the instantaneous approximation $Q_0=0$.

 Notice that the effective  kernel above was taken with 
a pure dependence on the momentum transfer $Q$. But, of course, the dependence on the quark and antiquark 
momenta could have been more complicated. The vector kernel   $J_v(Q)$     above would 
correspond to the one gluon exchange (with $P_{\mu\nu}$ depending on the adopted gauge)
while the scalar kernel $J_s(Q)$ 
would account for the nonperturbative interaction.  Taking $J_v \! = \!\displaystyle-{1\over 2 \pi^2} {4\over 3} 
{1\over {\bf Q}^2}$ and $J_s \! = \! \displaystyle -{\sigma\over \pi^2} {1\over {\bf Q}^4}$,  
      reproduce the Cornell potential and the
corrections in the nonrelativistic reduction of such kernel in the instantaneous approximation would give a form for the
 $V_{\rm SD} $ and the $V_{\rm VD}$.
The confining part of the kernel was usually chosen to be a Lorentz scalar in order to match the data 
on the fine separation on the $P$ states \cite{Schnitzer:1978gq}.
If one however takes a kernel which is not a pure convolution kernel, as it is to be expected in interactions
generated at higher orders, and deals more appropriately with the instantaneous approximation, one can get quite different
relativistic corrections, especially for the velocity dependent part \cite{Brambilla:1992fx}.
We conclude that this type of description is model dependent and does not allow for further progress. In particular,
the parameters of the model cannot be related to the underlying field theory and the systematics of the model cannot be
estimated.

\begin{figure}[ht]
\centering
\includegraphics[width=0.9\linewidth]{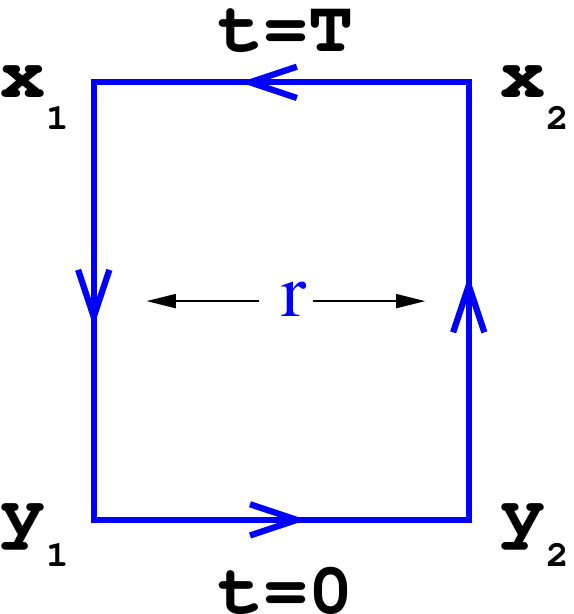}
\caption{The static Wilson loop. It contains the interaction of a static quark-antiquark pair created at a time $t=0$ 
and annihilated at a subsequent large time $T$. Initial and final states are made gauge invariant by the present 
of Schwinger line given in eq. (\ref{string}).}
\label{wilson}
\end{figure}

A more systematic procedure to  relate the potential to QCD has been  based
on the so called Wilson loop approach.
In order to obtain information on the structure of nonperturbative corrections to the potential it is  useful
to define the potential in terms of gauge invariant objects suitable for a direct lattice QCD evaluation.
Indeed lattice QCD is one of the best methods  to  extract nonperturbative information from QCD, for some reviews see
\cite{Kronfeld:2012uk,Rothe:1992nt,Kogut:1982ds,Bali:2000gf,Karsch:2001vs}.
Let us see how this works in  the case of the static potential.
Let us consider a locally gauge invariant quark-antiquark
color singlet state (for more details see \cite{Brambilla:1999ja,Lucha:1991vn}:

\begin{equation}
\vert \phi_{\alpha \beta}^{lj} \rangle \equiv {\delta_{lj}\over \sqrt{3}}
\bar{\psi}^i_\alpha(x) U^{ik}(x,y,C) \psi^k_\beta(y) \vert 0\rangle
\label{statgaug}
\end{equation}
where $i,j,k,l$ are colour indices (that will be suppressed in the following), 
$\vert 0\rangle$ denotes the QCD ground state and the Schwinger string line has the form
\begin{equation}
U(x,y;C) =P \exp \left\{ i g \int_y^x A_\mu(z) \, dz^\mu \right\} \,  ,
\label{string}
\end{equation}
where   $A_\mu= A_\mu^a, a=1,8$ is  the gluon vector potential of QCD in the fundamental representation, $g$  the QCD coupling constant, and the integral is 
extended along the path $C$. The operator $P$ denotes the path-ordering prescription
which is  necessary due to the  fact that $A_\mu$ are non-commuting matrices.
Let us see how
the quark-antiquark potential can be extracted from the quark-antiquark Green function constructed with such color singlet
states:
\begin{equation}
G(T) = \langle \phi({\bf x},0) \vert \phi({\bf y}, T)\rangle =
\langle \phi({\bf x},0) \vert \exp{(-iH T)}\vert \phi({\bf y}, 0)\rangle . 
\label{green}
\end{equation}
Inserting a complete set of energy eigenstates $\psi_n$ with eigenvalues $E_n$ and making a 
Wick rotation we find
\begin{eqnarray}
G(-i T) & = & \sum_n \langle \phi({\bf x},0) \vert \psi_n \rangle \langle \psi_n \vert 
 \phi({\bf y}, 0)\rangle \exp{(-E_n T)} \nonumber \\
&\to &   \langle \phi({\bf x},0) \vert \psi_0 \rangle \langle \psi_0 \vert 
 \phi({\bf y}, 0)\rangle \exp{(-E_0 T)} \quad  {\rm for }\quad T\to \infty
\label{inf}
\end{eqnarray}
which gives the Feynman--Kac formula for the ground state energy
\begin{equation}
E_0 = - \lim_{T \to \infty} {\log G(-iT) \over T}.
\label{fey}
\end{equation}
The only condition for the validity of Eq. (\ref{fey}) is that the $\phi$ states have a non-vanishing 
component over the ground state.
This is precisely the way in which  hadron masses are computed on 
the lattice. 
Of course, to maximize the overlap with the ground state in consideration appropriate operators may be used.
If the $\phi$ state denotes a state of two exactly static particles interacting at a distance $r$, then 
the ground state energy is a function of the particle separation, $E_0\equiv E_0(r)$, 
and gives the potential of the first adiabatic surface. It is possible to obtain
an explicit analytic  form  of the quark-antiquark Green function for infinitely heavy 
quarks ($m\to \infty$, static limit) and for large temporal intervals ($T\to \infty$),  see  \cite{Brambilla:1999ja}.
We consider  that at a time $t=0$ a quark and an antiquark pair is  created and  that they interact 
while  propagating  for a time $t=T$ at which they are annihilated. 
Then ($x_j=({\bf x}_j, T), y_j=({\bf y}_j, 0)$, see Fig. \ref{wilson}) we  obtain
\begin{eqnarray}
G_{\beta_1\beta_2\alpha_1\alpha_2}(T) &{\buildrel {m\to\infty}\over \longrightarrow }& \,
\delta^3({\bf x}_1-{\bf y}_1) \delta^3({\bf x}_2-{\bf y}_2) (P_+)_{\beta_1\alpha_1} (P_-)_{\alpha_2\beta_2}
\nonumber \\
& & \times e^{-2m T}\langle {\rm Tr}\, {\rm P}  e^{ig \oint_{\Gamma_0} dz_\mu A_\mu (z) }\rangle
\label{solgreen}
\end{eqnarray}
with $P_\pm\equiv  (1\pm \gamma_4)/2$. The integral in Eq. (\ref{solgreen}) extends over the circuit 
$\Gamma_0$ which is a closed rectangular path with spatial and temporal extension $r=\vert {\bf x}_1 -{\bf x_2}\vert$
and $T$ respectively, and has been formed by the combination of the path-ordered exponentials
along the horizontal (=time fixed) lines, coming from the Schwinger strings, and those along the 
vertical lines coming from the static propagators (see Fig. \ref{wilson}). The brackets in (\ref{solgreen}) denote 
the QCD vacuum expectation value, which  in Euclidean space is
\begin{equation}
\langle f[A]\rangle  \equiv {1\over Z} \int {\cal D} A f [A] e^{-\int d^4 x L^E_{YM}}.
\label{expece}
\end{equation}

From Eq. (\ref{solgreen}) it is clear that the dynamics of the quark-antiquark interaction is contained in 
\begin{equation}
W(\Gamma_0) =  {\rm Tr\, P}  e^{\displaystyle i g \oint_{\Gamma_0} dz_\mu A_\mu (z) }.
\label{wilsstat}
\end{equation}
This is the famous  static Wilson loop \cite{Wilson:1974sk}. 
In the limit of infinite quark mass considered, the kinetic energies of the quarks drop out of the theory, 
the quark Hamiltonian becomes identical with the potential  while the full Hamiltonian contains also 
all types of gluonic excitations. According to the Feynman--Kac formula the limit $T\to \infty $ projects
out the lowest state i.e. the one with the ``glue'' in the ground state. This  has the role of the quark-antiquark 
potential for pure mesonic states. We will see in the section over exotics that the excited states have the role
of the potential in the case of hybrid states, i.e. states with gluons contributing to the binding.
Now, comparing Eq. (\ref{solgreen}) with Eq. (\ref{inf}) and considering that 
the exponential factor $\exp{(- 2m T)}$ just accounts for the the fact that the energy of the 
quark-antiquark system includes the rest mass of the pair, we obtain
\begin{equation}
V_0(r) \equiv E_0(r) = -\lim_{T\to \infty} {1\over T } \log \langle W(\Gamma_0) \rangle.
\label{potfond}
\end{equation}
The heavy quark degrees of freedom have now completely disappeared and the expectation value in (\ref{potfond}) 
can  be evaluated in  the pure Yang--Mills theory   or in unquenched QCD (i.e. taking into account sea light quarks)
\cite{Wilson:1974sk,Brown:1979ya}.
Notice that the potential is given purely in terms of a gauge invariant quantity (the Wilson loop 
precisely). In this way we have reduced the calculation of the static potential to a well posed problem 
in field theory: to obtain the actual form of $V_0$ we need to calculate the QCD expectation value of the  static 
Wilson loop. The static Wilson loop is the low energy domain of QCD
is dominated by an area law, i.e. can be approximated at leading order at large $r$
as  $  \langle W(\Gamma_0) \rangle \sim  exp \{ -\sigma r T\} $ (i.e. the area of the loop in the exponent
multiplied by  the string tension), as shown in the strong coupling expansion
\cite{Rothe:1992nt,Creutz:1976ch,Bali:2000gf,Brambilla:1999ja,Creutz:1980zw}.
In the high energy
domain the first perturbative contribution comes from the one gluon exchange and therefore Eq. (\ref{potfond})
would reproduce  in these limits the Cornell potential of Eq. (\ref{cornell}), which appears to be  a simple superposition
of these two behaviours. We will come back to the area law behaviour of the Wilson loop in the section
on studies of confinement.
The Wilson loop  formalism have been used to obtain the form of spin dependent $V_{SD}$  potential in 
the famous papers of Eichten, Feinberg  \cite{Eichten:1980mw}  and Gromes \cite{Gromes:1984ma} and later
to obtain the form of spin independent, momentum dependent $V_{VD}$  potential in  the classical works 
\cite{Barchielli:1986zs,Barchielli:1988zp}. These works express the relativistic corrections  to the potential in terms of
generalized static Wilson loops containing field strengths (i.e. chromoelectric and chromomagnetic fields) insertion in
the quark lines. In \cite{Brambilla:1993zw} a nonstatic Wilson loop has been used to obtain the relativistic corrections
to the static three-quarks potential.

Up to the establishment of the NREFT called potential nonrelativistic QCD in the nineties
\cite{Brambilla:1999qa,Brambilla:1999xf,Brambilla:2004jw}, the Wilson  loop approach has been
the best theory founded method and was used
to calculate the potentials on the lattice  (see e.g. the review \cite{Bali:2000gf}).

However, this method was providing only part of the  QCD result. The Wilson loop relativistic
potential corrections  were missing all the non analytic term in the mass,  $\ln{m/\mu}$ terms, that were instead obtained
in a direct  purely perturbative QCD one loop calculation 
\cite{Gupta:1982kp}. Moreover, QCD in the static quark antiquark configuration  can still change colour by emission 
of gluons. This introduces a new dynamical scale in the evaluation of the potential
that should be taken into account as it became evident in the seminal paper
of Appelquist Dine and Muzinich \cite{Appelquist:1977tw} that identified infrared divergences in the three loops  perturbative fixed order
calculation of the static Wilson.
Moreover, already in \cite{Lucha:1991vn} the question was raised about why potential models where working so well.
It took a while and the development of non relativistic effective field theories
to answer this question and to be able to calculate systematically the potentials from  QCD.

All these problems are addressed and  solved in pNRQCD.

\section{\textit{Nonrelativistic effective field theories}}

As we mentioned, the study of quarkonium in the last few decades has witnessed two major developments:
the establishment of nonrelativistic effective field theories (NREFTs)  and  progress in lattice QCD calculations of excited states and resonances, with calculations at physical light-quark masses.
Both allow for precise and systematically improvable computations that are largely model-independent.
It is  precisely this advancement in the understanding of quarkonium and quarkonium-like systems  inside QCD
that makes today quarkonium exotics as  particularly valuable (see the section on Exotics).
In fact, today that we are confronted with a huge amount of high-quality data, which
have provided for the first time uncontroversial evidence for the existence of exotic hadrons,
by  using modern theoretical tools that allow us to explore in a controlled way these new forms of matter
we can get a  unique insight into the low-energy dynamics of QCD.

\begin{figure}[ht]
\centering
\includegraphics[width=0.6\linewidth]{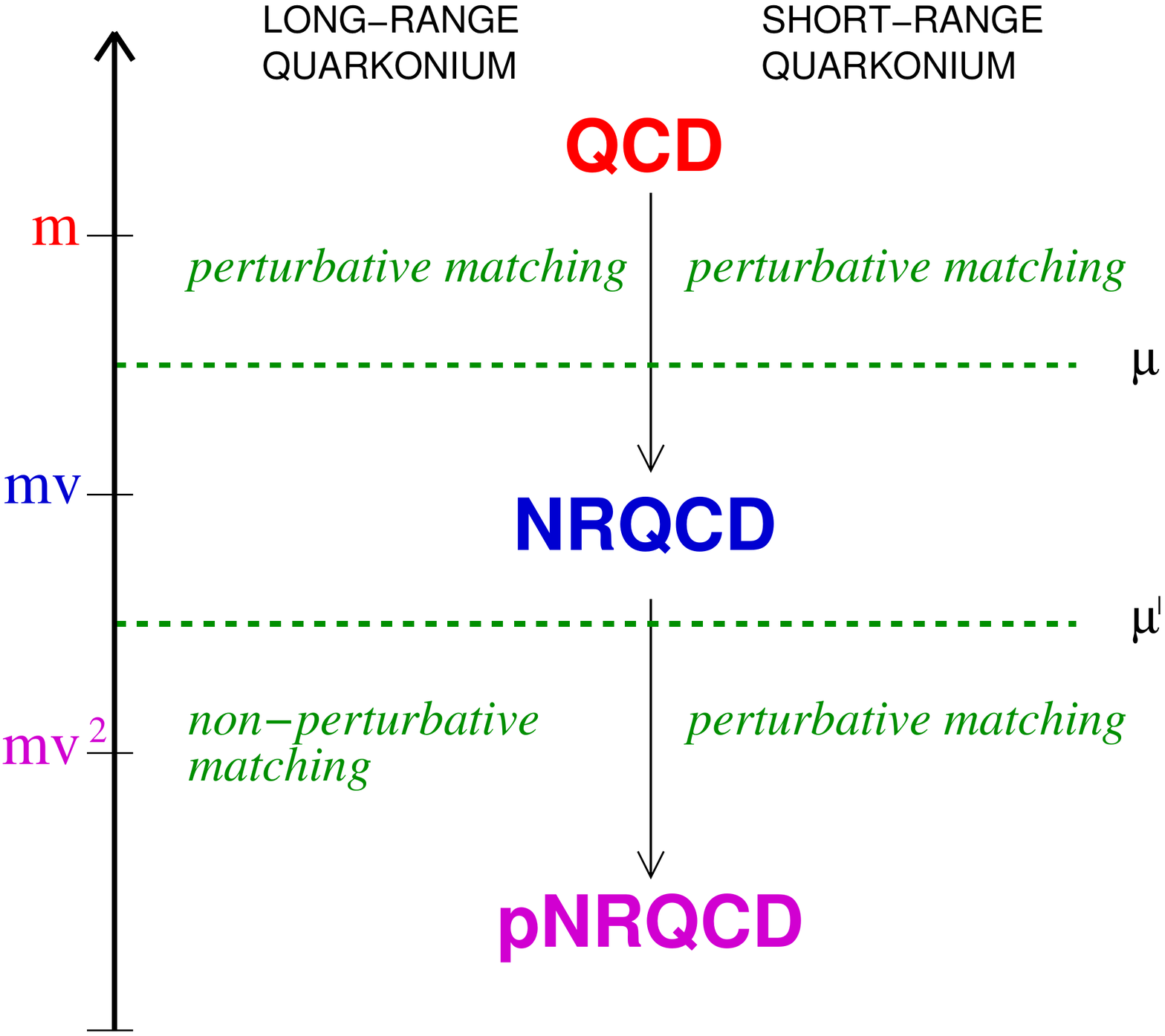}
\caption{The hard, soft and ultrasoft scales of quarkonium and the corresponding NREFTs that can be constructed. NRQCD follows from the integration 
of the gluons and  quarks at the hard scale, pNRQCD follows from integration of gluons at the soft scale. If the soft scale is bigger than 
$\Lambda_{\rm QCD} $ then the matching from NRQCD to pNRQCD is perturbative, otherwise it is nonperturbative. }
\label{scales}
\end{figure}

The appearance of a hierarchy of scales calls for the application of effective field theory (EFT) 
methods. However, ``heavy quark effective theory'' (HQET) \cite{Neubert:2005mu,Georgi:1991mr},
the EFT description of heavy-light mesons,
where only an ultraviolet mass 
scale $m$ and an infrared mass scale $\lQ$ appear, 
is not suitable for the description of heavy quarkonia, since HQET is 
unable to describe the dynamics of the binding.
The existence of several physical scales 
makes the theoretical description of 
quarkonium physics more complicated. All  scales get entangled in a typical 
amplitude involving a quarkonium observable. For example, quarkonium annihilation 
and production take place  at the scale $m$,  
quarkonium binding takes place at the scale
$mv$, which is the typical momentum exchanged inside the bound state, while 
very low-energy gluons and light quarks (also called ultrasoft  (US) degrees of freedom)  
live long enough that a bound state has time to form and,  therefore, are sensitive to the 
scale $mv^2$. Ultrasoft gluons are responsible for phenomena similar 
to the  the Lamb shift in  hydrogen atom.
This pattern  of  scales has  made  the description  in
Quantum  Field  Theory particularly  difficult.
The solution is to take advantage of the existence of the different  energy scales   
to substitute QCD with simpler but equivalent NREFTs.
A hierarchy of  NREFTs, see Fig.\ref{scales},  may be constructed by systematically integrating out 
modes associated with high-energy scales 
not relevant for the quarkonium system.
Such integration is made in a matching procedure that 
enforces the equivalence between QCD and the EFT at a given 
order of the expansion in $v$.
The EFT Lagrangian is factorized in matching coefficients,
encoding the high energy degrees of freedom  and low energy operators.
The relativistic invariance is realized via exact relations 
among the matching coefficients
~\cite{Manohar:1997qy,Brambilla:2003nt,Heinonen:2012km,Berwein:2018fos}
The EFT displays a power counting in the small parameter $v$, therefore we are able 
to attach a definite power of $v$ to the contribution of each EFT operators to 
the physical observables.
We will introduce in the next sections
nonrelativistic  QCD (NRQCD) \cite{Bodwin:1994jh,Caswell:1985ui,Lepage:1992tx}     and potential nonrelativistic QCD (pNRQCD)
\cite{Pineda:1997bj,Brambilla:1999xf}, see the review  \cite{Brambilla:2004jw}.

It turns to be instrumental to combine NREFTs and lattice.
In fact, on one hand, the NREFT is pulling out scales from observables
in a controlled way, separating them, and delivering  results for calculations at large scales including the logs resummation.                                    
On the other hand, the low energy contributions that the EFT has factorized are often nonperturbative and can be evaluated on the lattice.
This  greatly reduces the complexity of the problem allowing the lattice to target the nonperturbative part
directly, appropriately defined in the EFT in terms of gauge invariant, purely gluonic  objects, a big simplification with
respect to an ab initio lattice calculation of an observable. The latter is much more difficult because
still contains all the physical scales of the problem.
In this framework  also quenched lattice calculations can be pretty useful because at least at higher energy
the flavor dependence is accounted for by the EFT  matching coefficients.
We founded the TUMQCD lattice collaboration precisely to address these calculations directly inside the EFT \cite{tumqcd}.

\section{\textit{Nonrelativistic QCD}}

Nonrelativistic QCD (NRQCD)~\cite{Caswell:1985ui,Bodwin:1994jh}, 
follows from QCD  integrating out the scale $m$. As a consequence, 
the effective Lagrangian is organized as an 
expansion in $1/m$  and $\als(m)$: 

\begin{equation} \label{EQ:NRQCD}
{\cal L}_{\rm NRQCD}  = \sum_n \frac{c_n(\als(m),\mu)}{m^{n} } 
\times  O_n(\mu,mv,mv^2,...),
\end{equation}
where $c_n$ are Wilson coefficients that contain the 
contributions from the scale $m$. They  
can be calculated  via a well defined procedure call matching, see e.g.  \cite{Brambilla:2004jw},
in which one calculates  in perturbation theory Green Functions or amplitudes in QCD and in NRQCD, expand 
in $1/m$ and  insert the difference between the two calculations (i.e. the non analytic terms in the scale 
that is been integrated out, the mass) inisde the matching coefficients of the EFT.  In fact since in NRQCD 
we expand in the mass no contributions of the form $\log(m/\mu)$ can be  generated in calculations in the EFT.

The  $O_n$ are  local operators of NRQCD; the matrix 
elements of these operators contain the
physics of scales below $m$, in particular of the scales  $mv$
and $mv^2$ and also of the nonperturbative scale $\lQ$.
 Finally, 
the parameter $\mu$ is the NRQCD factorization 
scale. The low-energy operators $O_n$  are constructed out of two or four  
heavy quark/antiquark fields plus gluons.
At the lowest order in  the  $1/m$ expansion
the NRQCD Lagrangian density is
\begin{eqnarray}
{\cal L}_{\rm NRQCD} \! &=& \! \psi^\dagger \! \left\{ i D^0 + {{\bf D}^2\over 2 m} \right\} \!\psi +
\chi^\dagger \! \left\{ i D^0 - {{\bf D}^2\over 2 m} \right\} \!\chi \nonumber \\
& & - {1\over 4} F_{\mu \nu}^{a} F^{\mu \nu \, a}\,,
\label{lagnrqcd}
\end{eqnarray}
where $\psi$ is the Pauli spinor field that annihilates the fermion and $\chi$
is the Pauli spinor field that creates the antifermion; $i D^0=i\partial_0 -gA^0$ 
and $i{\bf D}=i{\bfnabla}+g{\bf A}$.  
At order $1/m^2$  bilinear terms in the quark (antiquark) field containing the chhromoelectric  and chromomagnetic fields, as well covariant derivatives and 
soin start to appear. These are of the same form that one would obtain from a Foldy-Wouthuysen transformation but in addition are multiplied 
by matching coefficients that contain the UV behaviour of QCD. In addition,
four fermion terms start to appear  at order   $1/m^2$ 
with matching coefficients containing both real and imaginary parts. The presence of imaginary parts describe the 
decays of quarkonium via annihilation of hard gluons that have been integrated out from the theory.

The quarkonium state $ \vert H\rangle$ in 
NRQCD is expanded in the number of partons
\begin{equation}
| H \rangle = | \bar{Q} Q  \rangle + | \bar{Q} Q  g \rangle + | \bar{Q} Q  \bar{q} q \rangle + \cdots  
\end{equation}  
where the states including one or more light parton are shown to be suppressed by powers of $v$.
In the  $| \bar{Q} Q  g \rangle$ for example the quark-antiquark are in a color octet state.
The NRQCD lagrangian has been extensively used on the lattice 
to calculate quarkonium spectra and decays.
On the other hand,  NRQCD has been deeply impactful on the study of quarkonium production at the LHC putting forward a factorization formula for
the inclusive cross section for the direct production of the
quarkonium $H$  at large transverse momentum
 written as a sum of products of NRQCD matrix elements and 
short-distance coefficients:
\begin{equation} \label{EQ:Prod} 
\sigma[H]=\sum_n \sigma_n \langle {\cal K}_n^{4 {\rm fermions}} \rangle
\end{equation}
where  the
$\sigma_n$ are short-distance coefficients, and the matrix elements
$\langle {\cal K}_n^{4 {\rm fermions}} \rangle$ 
are vacuum-expectation values containing  four-fermions operators in color singlet and color octet configurations identified by
  $Q \bar{Q} $angular momentum state quantum numbers.
  They contain in the middle a projector over the $Q\bar{Q} $ pair plus anything and some Schwinger lines with a particular path 
  to make them gauge invariant. The factorization for production  is proved only at order NNLO and in some cases
  \cite{Nayak:2005rt,Bodwin:2019bpf}.
  It is different from the NRQCD factorization of inclusive decays  that is proven at all orders. In such case the low energy part contains quarkonium expectation values 
  of color singlet and color octet four quark operators without the projector in the middle.
The matrix elements (LDMEs, long distance matrix elements)  $\langle {\cal K}_n^{4 {\rm fermions}} \rangle $
contain all of the nonperturbative physics associated
with the  evolution of the $Q\bar Q$ pair into a quarkonium state and they are universal. Their NRQCD
definition does not  allow a lattice calculation and their extractions from collider data is still
challenging \cite{Brambilla:2010cs,Chapon:2020heu,Andronic:2015wma,Bodwin:2013nua,Chung:2020gyb}.
Differently from HQET, the power counting of NRQCD is not unique, due to the fact that many physical scales are still 
dynamical ($mv, mv^2, \Lambda_{\rm QCD}$). This still complicates bound state calculations as the soft and US scale can get entangled. 

\section{\textit{ potential Nonrelativistic QCD}}

Quarkonium formation happens at the scale $mv$. 
The suitable NREFT is potential Nonrelativistic QCD,  
pNRQCD \cite{Pineda:1997bj,Brambilla:1999xf,Brambilla:2004jw},  which 
follows from NRQCD by integrating out the scale $mv \sim r^{-1} $.
The pNRQCD description  directly
addresses the bound state dynamics, implements the Sch\"odinger equation
as zero order problem, properly defines the potentials as matching coefficients,
and allows to systematically calculate relativistic and retardation corrections.
In this way even quantum mechanics can also  be reinterpreted as a pNREFT.
Moreover, since the scale $mv$ has been integrated out, the power counting of pNRQCD is less ambiguous 
than the one of NRQCD.

The  soft scale, proportional to the inverse quarkonium radius $r$,
may be either perturbative ($mv \gg \lQ$) or  nonperturbative  
($mv \sim \lQ$) depending on the physical system under consideration. 
We do not have any direct
information on the radius of the quarkonia, and thus the
attribution of  some of the lowest bottomonia and charmonia states 
to the perturbative or the nonperturbative soft regime may be  ambiguous, but it is likely that the lowest bottomonium and possibly also the 
lowest charmonium states have small enough radii that the scale $mv$
is in fact still perturbative. In a case with $\lQ \ll mv^2$ both scales are still perturbative and 
the system would be similar to a  Coulombic system, 
For such a quarkonium we would have  - similar as for the hydrogen atom -   
$v \sim \alpha_s (mv)$. However, none of the $b\bar{b}$ and $c\bar{c}$ states satisfy 
this condition. In all realistic quarkonia ($b\bar{b}$ and $c\bar{c}$) the ultrasoft scale is nonperturbative. 
Only for $t\bar{t}$ threshold states  the ultrasoft
scale  may  be considered still perturbative. 
In Fig.~\ref{scales} we schematically show the various scales. The short-range quarkonia 
are small enough to allow for a perturbative treatment of the scale $mv$, while for the long range 
quarkonia already this scale requires a non perturbative treatment. 

In the next section we will present weakly coupled and strongly coupled pNRQCD, discuss the form of the potential
and briefly address the countless  phenomenological applications of this that has become the standard treatment
for quarkonium. 
On one hand, pNRQCD allows us systematic and  precise determinations of processes at
high energy collider experiments and the definition and  computation
of quantities of large phenomenological interest, on the other hand the systematic factorization enables
us for studies of confinement.
 The  low energy nonperturbative factorized effects depend on the size of the physical system: 
when $\lQ$ is  comparable or smaller than $mv$ they are carried by correlators of 
chromoelectric  or chromomagnetic fields nonocal or local in time, when $\lQ$ is of order $mv$ they are carried 
by generalized static Wilson loops, non local in space, with insertion on chromoelectric and chromomagnetic fields.
The EFT allows us to make model independent predictions and we can use the power counting to attach an error to the 
theoretical predictions.  The nonperturbative physics in pNRQCD is encoded in few low energy correlators that depend only on the 
  glue and are gauge invariant: these are objects in principle ideal for lattice calculations.
 Even more interesting is that this EFT description allows modifications to be used to describe
exotics states (BOEFT) and the nonequilibrium evolution of quarkonium in medium (pNRQCD at finite temperature
with open quantum system) as we will address in the last two sections.

\subsection{\textit{Weakly coupled pNRQCD}}
When $mv \gg \lQ$, we speak about weakly-coupled pNRQCD because 
the soft scale is perturbative and the matching from NRQCD to pNRQCD 
may be performed in perturbation theory.
The lowest levels of quarkonium, like $J/\psi$, $\Upsilon (1S),\Upsilon (2S) \dots$,
may be described by weakly coupled pNRQCD, while the radii of the excited states 
are larger and presumably need to be described 
by strongly coupled pNRQCD. All this is valid for states away from strong-decay threshold, 
i.e. the threshold for a decay into two heavy-light hadrons. 
The effective Lagrangian is organized as an expansion in $1/m$  and $\als(m)$, 
inherited from NRQCD, and an expansion in $r$ (multipole expansion) \cite{Brambilla:1999xf,Brambilla:2004jw}
\begin{equation}
 L_{\rm pNRQCD}^{\rm weak}  =    \int d^3R\,       \int d^3r\,  
\sum_n \sum_k \frac{c_n(\als(m),\mu)}{m^{n}}  
 V_{n,k}(r,\mu^\prime, \mu) \; r^{k}  
\times O_k(\mu^\prime,mv^2,...) ,
\end{equation}
where ${\bf R}$ is the center of mass position and 
 $O_k$ are the operators of pNRQCD.
 We do not show explicitly the part of the lagrangian  involving
 gluons and light quarks, as this part coincides with the QCD one. The matrix elements of the operators 
 above
depend on the low-energy scale
$mv^2$ and $\mu'$, where $\mu^\prime$ is the pNRQCD factorization scale.  The $V_{n,k}$
are the Wilson coefficients of pNRQCD that encode the contributions 
from the scale $r$ and are nonanalytic in $r$. The $c_n$ are the NRQCD matching coefficients 
as given in (\ref{EQ:NRQCD}).

The degrees of freedom, which are relevant below the soft scale, and which 
appear in the operators $O_k$, are $Q\overline{Q}$ states (a color-singlet $S$ and a color-octet $O=O_a T^a$  state,
depending on  ${\bf r}$ and ${\bf R}$)
and (ultrasoft) gluon fields, which are expanded in $r$ as well (multipole expanded and depending only on ${\bf R}$).
pNRQCD makes apparent that the correct zero order problem is the Schr\"odinger equation.
Looking at the equations of motion of pNRQCD, we may identify 
$V_{n,0}= V_n$ with the $1/m^n$ potentials that enter 
the Schr\"odinger equation and 
$V_{n,k\neq 0}$ with the couplings of the 
ultrasoft degrees of freedom, which
provide corrections to the Schr\"odinger equation.
These  nonpotential interactions, 
associated with the propagation of low-energy degrees of freedom
 start 
to contribute at NLO in the multipole expansion.
 They are typically related to nonperturbative effects and are carried 
by purely gluonic correlator local or nonlocal in time: they need to be calculated on  the lattice.

In particular at the NLO in the multipole expansion  at leading order in the expansion in $1/m$,
 the pNRQCD Lagrangian  density is 

\begin{eqnarray}
& & {\cal L}_{\rm pNRQCD}^{\rm weak} =
{\rm Tr} \Biggl\{ {\rm S}^\dagger \left( i\partial_0 - {{\bf p}^2\over m} 
- V_s(r) + \dots  \right) {\rm S} 
\nonumber \\
& &\qquad + {\rm O}^\dagger \left( iD_0 - {{\bf p}^2\over m} 
- V_o(r) + \dots  \right) {\rm O} \Biggr\}
\\
& &\qquad + g V_A ( r) {\rm Tr} \left\{  {\rm O}^\dagger {\bf r} \cdot {\bf E} \,{\rm S}
+ {\rm S}^\dagger {\bf r} \cdot {\bf E} \,{\rm O} \right\} 
+ g {V_B (r) \over 2} {\rm Tr} \left\{  {\rm O}^\dagger {\bf r} \cdot {\bf E} \, {\rm O} 
+ {\rm O}^\dagger {\rm O} {\bf r} \cdot {\bf E}  \right\},  \nonumber
\label{pnrqcd0}
\end{eqnarray}
where ${\bf R} \equiv ({\bf x}_1+{\bf x}_2)/2$, ${\rm S} = {\rm S}({\bf r},{\bf R},t)$ and 
${\rm O} = {\rm O}({\bf r},{\bf R},t)$ are the singlet and octet quark-antiquark composite nonrelativistic  fields respectively. 
All the gauge fields in Eq. (\ref {pnrqcd0}) are evaluated 
in ${\bf R}$ and $t$. In particular ${\bf E} \equiv {\bf E}({\bf R},t)$ and 
$iD_0 {\rm O} \equiv i \partial_0 {\rm O} - g [A_0({\bf R},t),{\rm O}]$. 
$V_s^0$ and $V_o^0$ are the singlet and octet heavy $Q$-$\bar Q$ static  potentials
Higher-order potentials in the $1/m$ expansion and the centre-of-mass kinetic 
term are not shown here and are indicated by the dots. At higher order in the multipole expansion 
and in the $1/m$ expansion more operators appears containing US chromomagnetic and chromoelectric
fields but they are suppressed in the power counting in $v$.
 We define 
\begin{eqnarray}
V_s^0(r) &\equiv&  - C_F {\alpha_{V_s}(r) \over r}, 
\label{defpot}\\ 
V_o^0(r) &\equiv&  \left({C_A\over 2} -C_F\right) {\alpha_{V_o}(r) \over r}
\nonumber
\end{eqnarray}
which shows that the singlet potential is attractive and the octet one is repulsive.
$V_A$ and $V_B$ are the matching coefficients associated in the Lagrangian (\ref{pnrqcd0}) 
to the leading corrections in the multipole expansion. Both the potentials and the coefficients 
$V_A$ and $V_B$ have to be determined by matching pNRQCD with NRQCD at a scale $\mu$ smaller 
than $m v$ and larger than the US scales. Since, in particular, $\mu$ is 
larger than $\Lambda_{\rm QCD}$ the matching can be done perturbatively. At the lowest order in the coupling 
constant we get $\alpha_{V_s} = \alpha_{V_o} = \alpha_{\rm s}$, $V_A=V_B=1$. 
In order to have the proper free-field normalization in the colour space we define 
\begin{equation}
{\rm S} \equiv { 1\!\!{\rm l}_c \over \sqrt{N_c}} S \quad \quad {\rm O} \equiv  { T^a \over \sqrt{T_F}}O^a, 
\label{norm}
\end{equation}
where $T_F=1/2, N_c=3$.  

\leftline{\bf The perturbative QCD potential}

\begin{figure}[ht]
\centering
\includegraphics[width=0.9\linewidth]{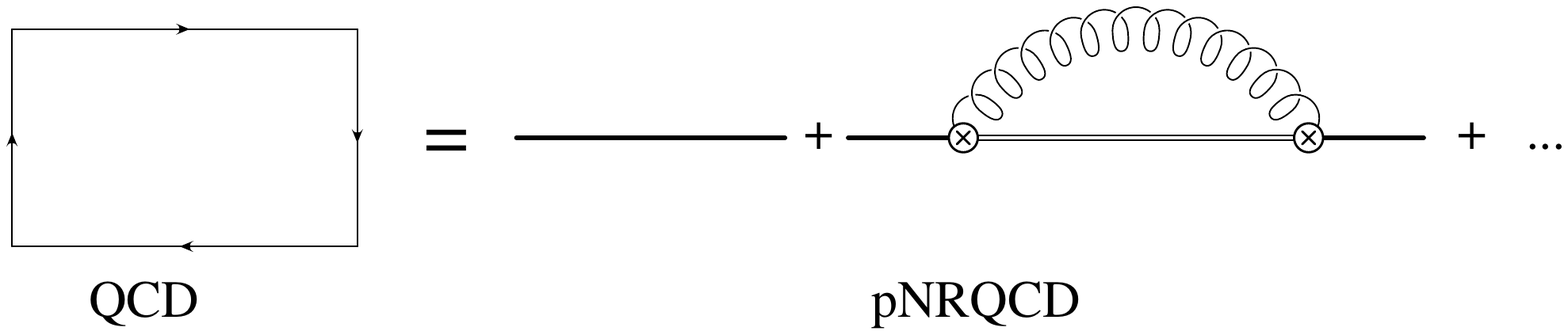}
\vspace{0.1cm}
\caption{The matching of the static potential. On the right side are the pNRQCD fields: 
simple lines are singlet propagator, double lines are octet propagators, 
circled-crosses  are the singlet-octet vertices of Eq. (\ref{pnrqcd0}) and 
the wavy line is the US gluon propagator.}
\label{figmat}
\end{figure}

For system with a perturbative soft scale
the potentials can be calculated in perturbation theory, i.e. no nonperturbative quantities enter the 
potential \cite{Brambilla:2004jw}. 
The potentials can be calculated at all order in the perturbative expansion in $\alpha_{\rm s}$
via a well defined matching procedure that entails to compare appropriate Green functions in NRQCD and in pNRQCD
calculated in perturbation theory up to the desired order of the perturbative expansion, order by order in  $1/m$ and order by order in the multipole expansion.
 The difference between the two is encoded 
in the matching coefficients. On the  pNRCD side lower energy scales may be expanded in the loop integral and give no contribution in the matching
in dimensional regularization.
The singlet static potential is calculated  by matching 
the relevant  NRQCD green function (the static Wilson loop) and the 
pNRQCD singlet  Green function see  Fig. \ref{figmat}).
At three loops, an ultrasoft  (US) divergence is emerging at fixed order calculation 
that can be regularized resumming series of diagrams containing the US scale which is the difference between the 
singlet and the octet potential,   see    \cite{Brambilla:1999qa,Brambilla:2004jw}.

This solves the problem raised by Appelquist Dine and Muzinich  \cite{Appelquist:1977tw},
 it explains how  this divergence cancels  between the potential ($\log r \mu$)
 and the contribution of the  US chomoelectric correlator  $\log ((V_o-V_S)/\mu )$  (second contribution 
 in the r.h. s. of Fig. (\ref{figmat}))  leaving a term 
 in $\log \alpha_s$ in the static energy  and qualifies the potential as a matching coefficient of an EFT which is depending on the US scale.
 Being matching coefficients of the effective field theory, the potentials 
 undergo renormalization, develop a scale dependence and satisfy renormalization 
 group equations, which allow  to resum large  US logarithms  see  e.g. \cite{Pineda:2000gza,Brambilla:2006wp,Brambilla:2009bi}.

 In particular one obtains the complete expression 
for the static potential, at 3 loops, up to the relative order $\alpha_{\rm s}^3 \ln \mu r$ 
in coordinate space ($\alpha_{\rm s}$ is in the $\overline{\rm MS}$ scheme) as:
\begin{eqnarray}
&&{\alpha}_{V_s}(r, \mu)=\alpha_{\rm s}(r)
\left\{1+\left(a_1+ 2 {\gamma_E \beta_0}\right) {\alpha_{\rm s}(r) \over 4\pi}\right.
\\
&&+\left[\gamma_E\left(4 a_1\beta_0+ 2{\beta_1}\right)+\left( {\pi^2 \over 3}+4 \gamma_E^2\right) 
{\beta_0^2}+a_2\right] {\alpha_{\rm s}^2(r) \over 16\,\pi^2}
\left. + {C_A^3 \over 12}{\alpha_{\rm s}^3(r) \over \pi} \ln{ r \mu}\right\}, \nonumber
\label{newpot}
\end{eqnarray}
where $\beta_n$ are the coefficients of the beta function, $a_1$
and $a_2$  are the constants at 1 and 2 loops respectively.
 We emphasize that this  US contribution 
to the static potential would be zero in QED.  From Eq. (\ref{newpot})  is clear 
that $\alpha_{V_s}$ depends on the US scale and as such is not 
 a short distance quantity.
 
 The evaluated terms clarify the long-standing issue of how the perturbative static potential 
should be defined at higher order in the perturbative series. In perturbation theory he static potential
does not coincide with the static Wilson loop starting from three loops.
 It should be emphasized that the separation between soft and US contributions is not an
artificial trick but a necessary procedure if one wants to use the static potential in a Schr\"odinger-like equation
in order to study the dynamics of $Q$-$\bar Q$  states of large but finite mass. In
that equation the kinetic term of the $Q$-$\bar Q$ system is US and so is the 
energy. Since the US gluons interact with the $Q$-$\bar Q$ system, their dynamics is sensitive
to the energies of the (non-static) system and hence it is not correct to include them in
the static potential. When calculating a physical observable the $\mu$ dependence  must
cancel against $\mu$-dependent contributions coming from the US gluons. 
Finally it is worth mentioning that the static potential suffers from IR 
renormalons ambiguities with the following structure
$$
\delta V_s \sim C + C_2 r^2 + .\,.\,.
$$
The constant $C$ is known to be cancelled by the IR pole mass renormalon 
 \cite{Brambilla:2004jw}.
 The static singlet potential is currently know at NNNLL \cite{Brambilla:2009bi}, the only unknown at 
four loops is the constant term.

The static energy in the perturbative regime  has the form
\be
E_0(r)= V_s^0(r;\mu ) + \delta_{\rm US}(r,V_s^0,V_o^0, \dots;\mu),
\label{e0}
\ee
where $ \delta_{\rm US}(r,V_s,V_o,V_A,V_B, \dots;\mu)$ ($\delta_{\rm US}$ for short)
contains contributions from the ultrasoft gluons.
As mentioned, $V_s(r;\mu )$ and $V_o(r;\mu )$ do not depend on $\mu$ up to N$^2$LO \cite{Brambilla:1999xf}. 
The former coincides with $E_0(r)$ at this order and the latter may be found in \cite{Kniehl:2004rk}. 
The fact that the $\mu$ dependence of $\delta_{\rm US}$ must cancel 
the one in $V_s(r;\mu )$ is the key observation that 
leads to a drastic simplification in the calculation of the $\log \als$
terms in $E_0(r)$. So, for instance, the logarithmic contribution at N$^3$LO, 
which is part of the three-loop contributions to $V_s(r;\mu)$, may be extracted from a one-loop
calculation of $\delta_{\rm US}$ \cite{Brambilla:1999qa,Brambilla:1999xf} and the single logarithmic contribution at N$^4$LO,
which is part of the four-loop contributions to $V_s(r;\mu)$,
may be extracted from a two-loop calculation of $\delta_{\rm US}$.
Notice that we denote N$^n$LO, contributions to the potential of order
$\als^{n+1}$ and N$^n$LL, contributions  of order $\als^{n+2} \log^{n-1}\als$.

The general form of the relativistic corrections to the singlet potential 
in the center of mass is (we drop the index $s$ for simplicity):
\bea
V(r)&=&V^{(0)}(r) +{V^{(1)}(r) \over m}+{V^{(2)} \over m^2}+\cdots,  
\label{ppot}
\\
V^{(2)}&=&V^{(2)}_{SD}+V^{(2)}_{SI},\nn\\
V^{(2)}_{SI}
&=&
{1 \over 2}\left\{{\bf p}^2,V_{{\bf p}^2}^{(2)}(r)\right\}
+{V_{{\bf L}^2}^{(2)}(r)\over r^2}{\bf L}^2 + V_r^{(2)}(r),
\\
V^{(2)}_{SD} &=&
V_{LS}^{(2)}(r){\bf L}\cdot{\bf S} + V_{S^2}^{(2)}(r){\bf S}^2
 + V_{{\bf S}_{12}}^{(2)}(r){\bf S}_{12}({\hat {\bf r}}), \nn
\eea
${\bf S}={\bf S}_1+{\bf S}_2$,  ${\bf L}={\bf r}\times {\bf p}$,
 ${\bf S}_1=\bfsigma_1/2$, ${\bf S}_2=\bfsigma_2/2$, 
  and ${\bf S}_{12}({\hat {\bf r}}) \equiv 
3 {\hat {\bf r}}\cdot \bfsigma_1 \,{\hat {\bf r}}\cdot \bfsigma_2 - \bfsigma_1\cdot \bfsigma_2$.

We see that differently from what discussed previously the corrections to the potential start at order $1/m$.
 The potential proportional to $V_{LS}^{(2)}$ may
be identified with the spin-orbit
potential, the potential proportional to $V_{S^2}^{(2)}$ with the spin-spin potential and the potential proportional to $V_{S_{12}}^{(2)}$
with the spin tensor potential.
The above potentials read at leading (non-vanishing) order in perturbation theory (see, e.g., Ref.~\cite{Brambilla:2004jw}):
\begin{eqnarray}
& 
V^{(1)}(r) = -\frac{2 \alpha_s^2}{r^2} \,,
\label{V1pNRQCD} \\
&
V_r^{(2)}(r) = \frac{4\pi}{3} \alpha_s \delta^{(3)}({\bf r}\,) \,,
\hspace*{0.90cm}
V_{p^2}^{(2)}(r) = -\frac{4\alpha_s}{3r} \,,
\hspace*{0.90cm}
V_{L^2}^{(2)}(r) = \frac{2\alpha_s}{3 r} \,,
\label{V2p2pNRQCD} \\
&
V_{LS}^{(2)}(r) = \frac{2\alpha_s}{r^3} \,,
\hspace*{1.00cm}
V_{S^2}^{(2)}(r) = \frac{16\pi \alpha_s}{9} \delta^{(3)}({\bf r}\,) \,,
\hspace*{1.00cm}
V_{S_{12}}^{(2)}(r) = \frac{\alpha_s}{3 r^3} \,.
\label{V2S12pNRQCD}
\end{eqnarray}
These potentials and  the static octet potential have been calculated at higher order in perturbation theory
see e.g. \cite{Brambilla:2004jw,Pineda:2011dg,Kniehl:2004rk}.
Once one has obtained the potentials as matching coefficients of pNRQCD one can start doing calculations inside the EFT 
obtaining for example the energy levels of quarkonium \cite{Brambilla:1999xj,Kniehl:2002br}. Nonperturbative effects in the form of local 
or time nonlocal chromoeletric correlators start to appear at order $\als^5$ i.e. at NNNLO \cite{Brambilla:2004jw,Pineda:2011dg}:
their contribution is suppressed. 

The case of baryons made by three heavy quarks or two heavy quarks and a  light quark  has also been addressed in pNRQCD  \cite{Brambilla:2005yk} 
and the static three quark potential has been calculated at one loop accuracy, showing some interesting features \cite{Brambilla:2009cd}.

\subsection{\textit{Strongly  coupled pNRQCD}}
\begin{figure}[!tbp]
 \begin{minipage}[b]{0.5\textwidth}
    \includegraphics[width=\textwidth]{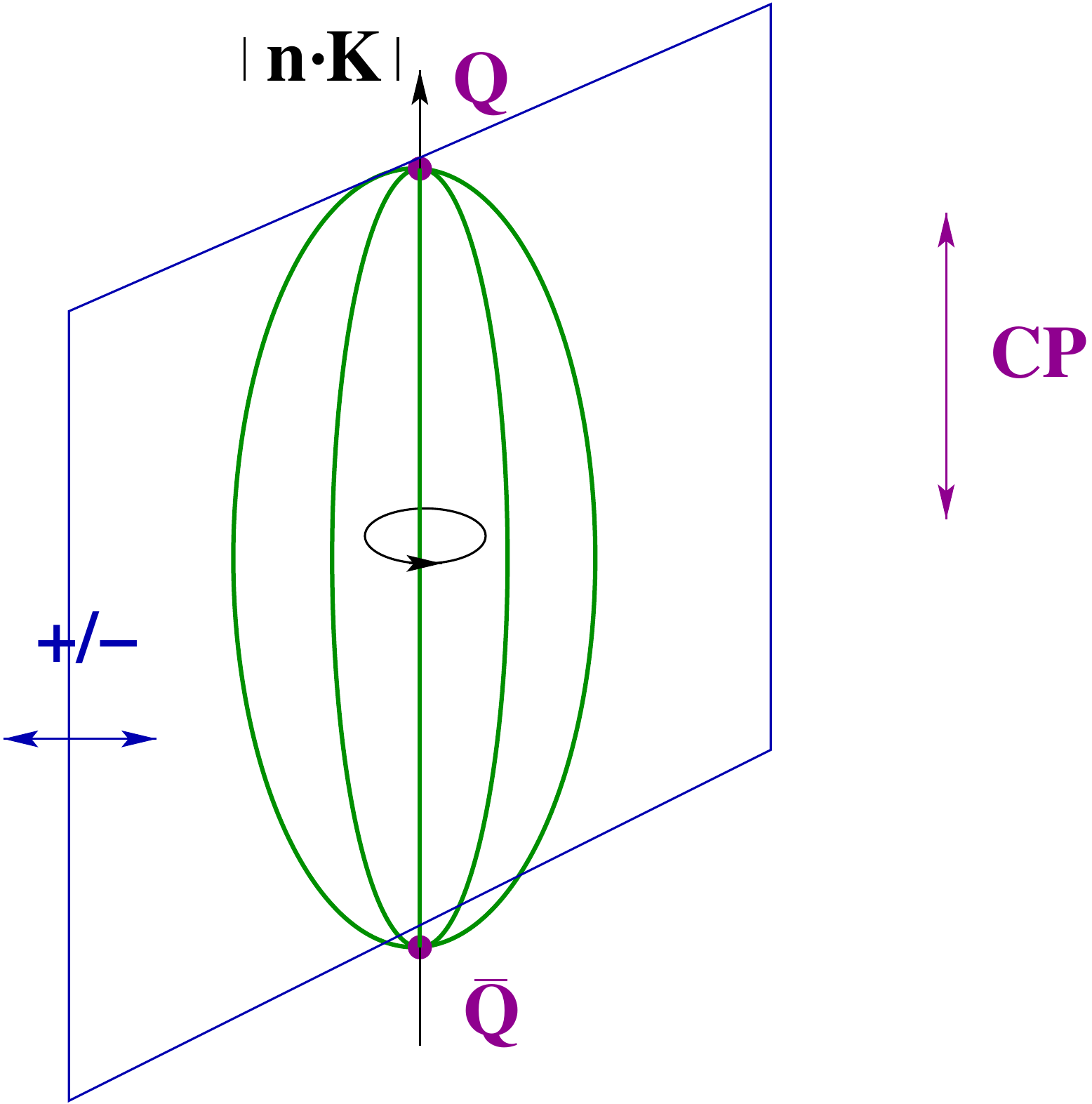}  
    \vskip 1.2 truecm
    \caption{Quarkonium hybrid symmetries.} \label{figsym}
     \end{minipage}
\hfill
  \begin{minipage}[b]{0.5\textwidth}
    \includegraphics[width=\textwidth]{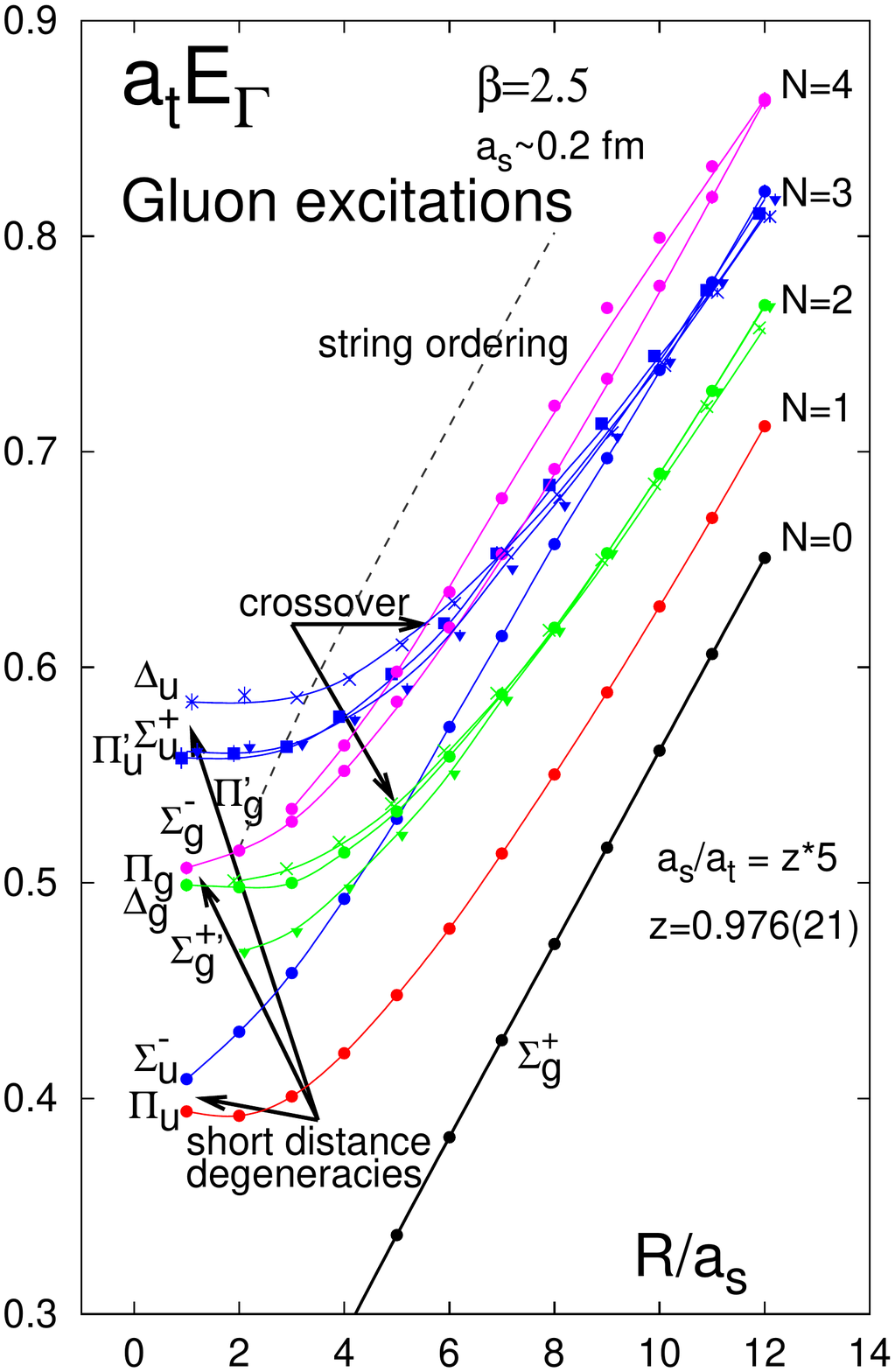} 
    \caption{Hybrid static energies, $E_\Gamma$, in lattice units, from~\cite{Juge:2002br}.}
      \label{figEg}
 \end{minipage}      
\end{figure}

When $mv  \sim \lQ$, we speak about  
strongly-coupled pNRQCD because the soft scale 
is nonperturbative and the matching 
from NRQCD to pNRQCD is nonperturbative i.e. it cannot be performed 
within an expansion in $\alpha_s$.
In this case, since we  work at a nonperturbative scale, only color singlet degrees of freedom remain dynamical and they include $Q\bar{Q}$
states, hybrids $Q\bar{Q} g$ states and glueballs  (and if we consider light quarks as part of the binding, color singlet states formed  by a heavy quark and heavy antiquark 
and light quarks, as in the case of tetraquarks).
  Since the physics is nonperturbative
we need to use  input coming from the lattice to construct the EFT.
In particular we can use  the lattice evaluation of the gluonic static energies of a  $Q\bar{Q}$ pair: they
have been calculated since
long \cite{Juge:2002br,Bali:2003jq}
and recently updated  \cite{Muller:2019joq}. Such lattice calculations  use insertions 
of gluonic fields at the initial and final Schwinger strings, inside a generalized static Wilson loop,   to select some  given symmetries.
  The gluonic static energies, $E_\Gamma$ in Fig.~\ref{figEg},
are classified according to representations of the symmetry group $D_{\infty\,h}$, typical of diatomic molecules, 
and labeled by $\Lambda_\eta^\sigma$ (see Fig.~\ref{figsym}):
$\Lambda$ is the rotational quantum number $|\hat{\bf r}\cdot{\bf K}| = 0,1,2,\dots$,
with ${\bf K}$ the angular momentum of the gluons, 
that corresponds to $\Lambda = \Sigma, \Pi, \Delta, \dots$;
$\eta$ is the CP eigenvalue ($+1\equiv g$ (gerade) and $-1 \equiv u$ (ungerade));
$\sigma$ is the eigenvalue of reflection with respect to a plane passing through the $Q\bar{Q}$ axis.
The quantum number $\sigma$ is relevant only for $\Sigma$ states.
In general there can be more than one state for each irreducible representation of $D_{\infty\,h}$: 
higher states are denoted by primes, e.g., $\Pi_u$, $\Pi_u'$, $\Pi_u'', \dots$
 Notice that this set of static energies will be fundamental later to address the exotics.

Since at this moment we are dealing with $Q\bar{Q}$ states below threshold we are interested merely in
the $\Sigma_g^+$ static energy (that in this case coincides with the static singlet $Q\bar{Q}$ potential)
and in the information that such curve develops a gap of order $\lQ$ at a distance $r\sim \lQ^{-1}$: therefore
all the hybrid static energies can be integrated out as pNRQCD follows from integrating out all degrees of freedom
with energy up to $mv^2$.
The quarkonium singlet field $S$ 
 is now the only low-energy dynamical 
degree of freedom in the pNRQCD Lagrangian (up to US pions)
which reads \cite{Brambilla:2000gk,Pineda:2000sz,Brambilla:2004jw}:
\begin{equation}
\quad   L_{\rm pNRQCD}=   \int d^3R\,       \int d^3r\, 
{ S}^\dagger
\left(i\partial_0-\frac{{\bf p}^2}{2m}-V_S(r)\right){S}\,
\label{strp}
 \end{equation}
 and lends support to potentials models in this particular regime.
 Indeed Eq. (\ref{strp}) originates a Sch\"odinger equation  governed by the singlet
 potential.
 
\leftline{\bf The nonperturbative QCD potential}
 
 The difference  with the phenomenological potential models
 is that now  the singlet potential $V_s(r) = V_0 + V^{(1)}/m +V^{(2)}/m^2$,  is the QCD potential
 calculated in the matching obtained in  \cite{Brambilla:2000gk,Pineda:2000sz}
 and  display  novel  characteristics. These potentials are nonperturbative and are given in terms of generalized
 Wilson loops.
However relevant differences emerge also with respect to the Wilson loop approach.
 The general decomposition of the potential is of the form of Eq. (\ref{ppot}). It features a contribution 
 already at order $1/m$ \cite{Brambilla:2000gk} given in terms of a novel expectation value of a static Wilson loop 
with insertion of a chromoelectric field.
 The potential at order $1/m^2$ appears factorized in the product of NRQCD matching coefficients, carrying  
 contribution in the $\log(m/\mu)$ and generalized Wilson loops with chromomagnetic and chromoelectric insertions.
 This solves the problem of the incompatibility of 
 the Wilson loop potentials and the perturbative calculation.  Moreover, additional generalized Wilson loops contribution of novel type emerge
 see \cite{Pineda:2000sz,Brambilla:2004jw} for all the explicit expressions.
 Spin effects  in quarkonium are appearing only at order $1/m^2$ and are
 therefore suppressed in the spectrum and in the transitions: we will see that things are different for exotics.
Some of these generalized Wilson loops  have been calculated on the  lattice
(only quenched) \cite{Bali:1997am,Bali:2000gf,Koma:2006si,Koma:2007jq}
but some contributions at order $1/m^2$ are still to be calculated.
The EFT then lends a clean definition and an interpretation
of the static Wilson loops measured on the lattice as actual potentials in this regime, together with a prescription to use them
to calculate observables.
Once the nonperturbative potentials are given in terms of generalized   Wilson loops one can use a model of
low energy QCD to evaluate them. A minimal area law for example gives for the potentials of Eq.(\ref{pott}) 
a linear term  for 
$V^{(0)}(r)= \sigma r$ and $   V_{L^2}^{(2)}(r) = -{\sigma\over 6} r$ and nonzero contributions to
$V_r^{(2)}(r)$ and $V_{LS}^{(2)}(r)$. The velocity dependent relativistic nonperturbative contribution corresponds
to the angular momentum and the energy of the flux tube between the quark and the antiquark while
the one in  the spin dependent part is relevant to obtain agreement with the
fine separations of the quarkonia multiplets (as it was pursued previously using a scalar confinement kernel in the Bethe-Salpeter equation).
These findings are in agreement with what is  obtained from the lattice  calculations of the Wilson loops.
Using a QCD effective string model add several further subleading corrections see \cite{Brambilla:2014eaa,Perez-Nadal:2008wtr}.

Using these potentials,  all the masses for heavy quarkonia below threshold
can be obtained by solving the Schr\"odinger equation
with such potentials. Lorentz invariance is still there in the form of exact relations among potentials and it has been observed on the lattice.
Decays are described by calculating the imaginary parts of the potentials \cite{Brambilla:2002nu} where nonperturbative contributions
enter in the form of gauge invariant time non local chromoelectric and chromomagnetic correlators that have still to be calculated on the lattice.
Summarizing, strongly coupled pNRQCD
factorizes low energy nonperturbative contributions in terms of generalised gauge invariant Wilson loops, opening the way
to a systematic study of the confinement mechanism and systematic applications to quarkonium spectrum and decay as we will see in the dedicated section.

The three heavy quark static potential has been studied  on the lattice using Wilson loops \cite{Takahashi:2000te,Takahashi:2002bw}.

\subsection{\textit{Spectra, transitions, decays and production, SM parameters extractions}}

Together with lattice, pNRQCD  is the theoretical framework that is nowadays  mostly used for  calculations  and predictions
of quarkonium properties. The power counting of the EFT allows to attach an error to each prediction.
In the regime in which the soft scale is perturbative, pNRQCD enables precise and systematic higher order calculations on bound state allowing the extraction of 
precise determinations of standard model parameters like the quark masses and $\alpha_s$.
For example, based on Eq. (\ref{e0}), it has been possible to use lattice calculations of the static energy with 2+1 flavour 
and the NNNLO pNRQCD calculation of the static energy, including the US log resummation, to extract a precise determination of 
$\als$  at rather low energy and run it at the mass of the Z, obtaining $\als(M_Z)= 0.11660^{+0.00110}_{-0.00056}$, which is a competitive extraction 
made at a pretty high orders of the perturbative expansion \cite{Bazavov:2019qoo,Brambilla:2009bi,Bazavov:2012ka}.
This method of $\als$ extraction is now used 
by several groups, see  e.g. \cite{Ayala:2020odx,Takaura:2018vcy} and  the force, defined in terms of a single chromolectric insertion in the Wilson loop
could be used as well \cite{Brambilla:2021wqs,Vairo:2016pxb}.

In the same way one can extract precise determinations of the bottom and charm masses using the experimental measurements of 
the mass of the lowest states and comparing it to the formula for the energies in pNRQCD at NNNLO, which depends on 
the mass in a given scheme. The renormalon ambiguity cancels between the mass and the static potential and a pretty good determination is possible
see e.g. \cite{Peset:2018ria, Kiyo:2014uca} and references therein.
Moreover, the energy levels of some of the lowest quarkonia states have been obtained  at high orders in perturbation theory using weakly coupled
pNRQCD  showing that the constant separation on the energy levels may be generated in this way, see  e.g. \cite{Brambilla:2001qk,Brambilla:2001qk,Brambilla:2000db,Peset:2015vvi}
Electromagnetic M1 and E1 transitions have been calculated in pNRQCD, see e.g. \cite{Brambilla:2005zw,Brambilla:2012be,Pineda:2013lta,Segovia:2018qzb}.
There are so many results that it is impossible to discuss  all of them here and we refer you to some reviews \cite{Pineda:2011dg,Brambilla:2004jw,Brambilla:2010cs,QuarkoniumWorkingGroup:2004kpm,Brambilla:2014jmp}.

For what concern production  it is very promising that by factorizing the quarkonium productions cross section at  lower energy in pNRQCD
\cite{Brambilla:2021abf, Brambilla:2020ojz,Brambilla:2020xod},
one can rewrite the octet NRQCD LDMEs, which are the nonperturbative unknowns, in terms of product 
of wave functions and  gauge invariant low energy correlators
that depend only on the glue and not the on flavor quantum numbers. This allows us to reduce by half 
the number of nonperturbative unknowns and  promises to have a great impact in the progress of the field.

\begin{figure}[ht]
\centering
\includegraphics[width=0.6\linewidth]{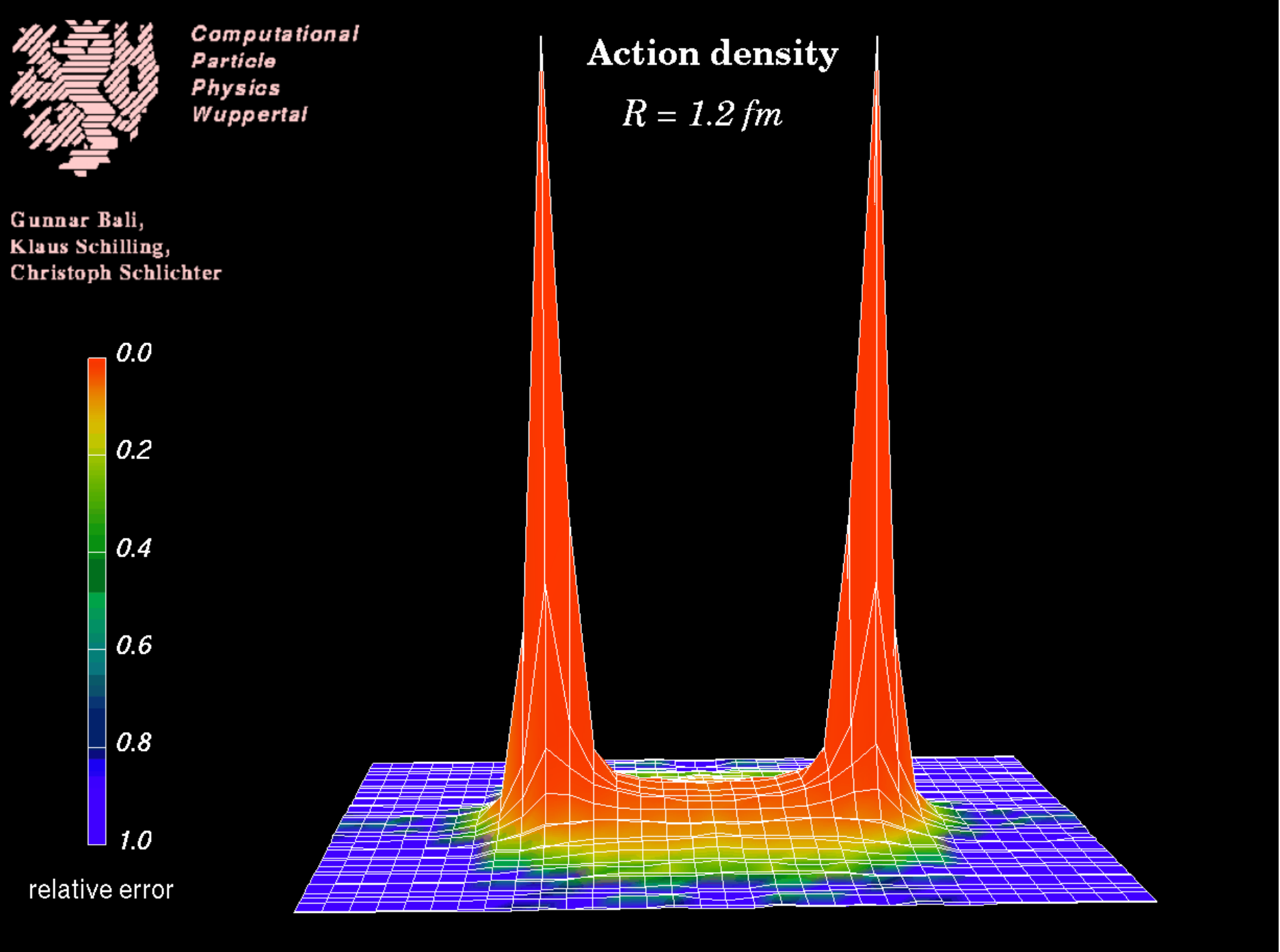}
\caption{The origin of  the linear potential between the static quark and antiquark may be traced back to a flux tube: a string
  of gluon energy between the quark pair. Here we present the hystorical picture of the action density distribution between
a static quark antiquark couple in $SU(2)$ at a physical distance of 1.2 fm, from \cite{Bali:1994de}.}
\label{flux}
\end{figure}

\section{\textit{Confinement and low energy QCD models}}

Strongly coupled pNRQCD realizes a  scale factorization  encoding the low energy physics 
in generalized gauge invariant Wilson loops, i.e. Wilson loops with insertions of chromoelectric and chromomagnetic fields.
Such objects no longer depends on the heavy quark degrees of freedom and on the  quark flavor.
It turns out that this is a systematic method to study the QCD confinement properties and 
put them directly in relation to the quarkonium phenomenology.
Indeed lattice QCD seems more suitable to ask 'what' instead of 'why' and to understand
the mechanism underlying confinement it may be useful to build models of low energy QCD and compare to the lattice results:
the interface is offered by the EFT.
We have seen that the area law emerging in the static Wilson loop at large distance is responsible of confinement, this in turn corresponds
to the formation of a chromolectric flux tube between the quark and the antiquark that sweeps the area of the Wilson loop, see Fig. \ref{flux}.
This effects is originated by the nonperturbative QCD vacuum that could be imagined as a disordered medium with whirlpools of colour 
on different scales, thus densely populated by fluctuating fields whose amplitude is so large that they cannot be described 
by perturbation theory. A QCD vacuum model can be established by making an assumption on the behaviour of the Wilson loop that gives the static potential. 
The relativistic corrections that involve insertions of gluonic fields in the Wilson loop follow then via functional derivative 
with respect to the  quark path see \cite{Brambilla:1993zw,Brambilla:1999ja}.
Then, one can notice for example that the part proportional to the square of the angular momentum in the of the $V_{vd}$ potential  obtained in strongly
coupled pNRQCD 
takes into account the energy and the angular momentum of the flux tube, which is something that could not be obtained 
in any Bethe Salpeter approach with a confinement scalar convolution kernel.
Lattice simulations of the  action density or the energy density between  the static quark and the static antiquark 
show indeed a  chromoelectrix  flux tube configuration see Fig.  \ref{flux} from ref. \cite{Bali:1994de} and
more recent calculations in \cite{Bicudo:2018jbb,Yanagihara:2019foh,Baker:2021jnr}.
The mechanism underlying confinement and flux tube formation has been investigated since long on the lattice 
\cite{Greensite:2003bk} using the Wilson loops and the 't Hooft abelian projection, to identify  the roles 
of magnetic monopoles  \cite{Amemiya:1998jz,Sasaki:1994sa} and center vortices, see e.g. the review \cite{Brambilla:2014jmp}. 

In the continuum several models of low energy QCD has been used to explain  the flux tube formation ranging  from
the dual Meissner effect and a dual abelian Higgs model picture, dual QCD \cite{Baker:1991bc}, the stochastic vacuum \cite{Dosch:1988ha},
the flux tube model \cite{Isgur:1984bm} and an effective   QCD string description.
  Each of these models can be used to obtain analytic estimates
  of the behaviour of the generalized Wilson loops for large distance, which in turn 
  give  the static potential and the relativistic corrections $V_1, V_{sd}, V_{vd}$  as function of $r$,
  see \cite{Baker:1996mk,Brambilla:1996aq,Baker:1998jw,Brambilla:2014eaa,Perez-Nadal:2008wtr}.
  Similar nonperturbative configurations leading to confinement can be studied
  analyzing the Wilson loop  in case of baryons with three or two heavy quarks 
\cite{Nawa:2006gv,Soto:2021cgk}.

\section{\textit{BOEFT and X Y Z exotics}}

Exotic states, i.e. states   with a composition different
from a quark-antiquark or three quarks in  a color singlet,
have been predicted before and after the advent of QCD.
In  the last decades  a large number of states, either with a manifest different composition (with isospin
and electric charge  different from zero) or with other exotic characteristics
have been observed  in the sector
with two heavy quarks $Q \bar{Q}$, at or above the quarkonium strong decay threshold at the
B-Factories, tau-charm and LHC  and Tevatron collider experiments
\cite{Brambilla:2019esw,Brambilla:2010cs,QuarkoniumWorkingGroup:2004kpm,Brambilla:2021mpo}. 
These states have been termed $X, Y, Z$  in the  discovery publications,
without any special criterion, apart
from  $Y$  being used for exotics with vector
 quantum numbers, i.e., $J^{PC} = 1^{--}$ . Meanwhile, the Particle Data Group (PDG) has proposed a new naming scheme
 \cite{ParticleDataGroup:2018ovx},  that extends the scheme used for ordinary quarkonia, in which
 the new names carry information on the $J^{PC}$ quantum numbers, see \cite{Brambilla:2019esw} for more details.
Some of these exotics have quantum numbers that cannot be obtained with ordinary hadrons. In this case, the
identification of these states as exotic is straightforward. In the other cases, the distinction requires a careful analysis
of experimental observations and theoretical predictions.
Of course all hadrons should be color singlets but allowing combinations beyond the $Q \bar{Q}$ and $QQQ$ in the two heavy quarks sector
calls for  tetraquarks like  $Q \bar{Q}  q \bar{q}$, $QQ  \bar{q} \bar{q}$,  pentaquarks  like $Q \bar{Q}  q q q$, hybrids $Q \bar{Q}  g$
and so on \cite{Ali:2017jda} .  We have observed these exotics up to now  only in the sector with two heavy quarks likely due to the fact
that the presence of the two heavy quarks stabilizes them. 
 Some of the discovered states have an  unprecedentely  small width even if they are at  or above the 
strong decay threshold.
$ X Y Z$ states  offer us unique possibilities  for the investigation of  the dynamical properties of
strongly correlated systems in QCD:  we should
develop the tools to gain a solid interpretation from the underlying field theory, QCD.                                      
This is a very significant problem with trade off to other fields featuring strong correlations and a pretty interesting     
connections to  heavy ion physics, as propagation of these states in medium may help us to scrutinize their properties.

Since  the new quarkonium revolution i.e. the discovery  of the first exotic state, the $X(3872)$  at BELLE in 2003 \cite{Belle:2003nnu},
a wealth of theoretical papers appeared  to  supply interpretation  and understanding of  the characteristics of the exotics.
Many models  are based  on the choice of  some dominant degrees  of freedom and an assumption on   the related interaction
hamiltonian. An effective field theory  molecular description of some of these states particularly
close to threshold was also put forward, see e.g.
\cite{AlFiky:2005jd,Guo:2017jvc,Braaten:2007dw,Braaten:2003he,Fleming:2011xa}. 
A priori the simplest system consisting of only two quarks and two antiquarks (generically called tetraquarks)
 is already a very complicated object and it is unclear whether or not any kind of clustering occurs in it.
 However, to simplify the problem it is common to focus on certain substructures and investigate their implications:
 in hadroquarkonia the heavy quark and antiquark form a compact core surrounded by a light-quark cloud; in compact tetraquarks the relevant
 degrees of freedom are compact
 diquarks and antidiquarks; in the molecular picture two color singlet mesons are interacting at some typical distance, for a review
 see \cite{Brambilla:2019esw}.
Discussions about exotics therefore often concentrate on the  'pictures' of the states,
like for example the   tetraquark  interpretation against the molecular one (of which both several different realizations exist).
However,  as a matter of fact all the light  degrees of freedom  (light quarks, glue, in different configurations)   should  be there in QCD
close or above the strong decay threshold: it  is a   result  of the strong dynamics which one sets  in
and  which configuration   dominates  in a given regime. 

Even in an ordinary quarkonium or in a heavy baryon, which has a dominant  $Q\bar{Q}$  or $QQQ$ configuration,  subleading contributions
of the quantum field theoretical Fock space may contribute,  with  have additional quark-antiquark pairs and active gluons.
However, in the most interesting region,  close or above the strong decay  threshold, where the X Y Z have been discovered,
 the situation is much more complicated: 
 there is no mass gap between quarkonium and the creation of a heavy-light mesons couple, nor to gluon 
 excitations, therefore  many additional states built on the light quark quantum numbers may appear \cite{Brambilla:2021mpo}.
Still, $m$  is a  large scale and   a first scale factorization is applicable so that Nonrelativistic QCD is still valid.
Then, if we want to introduce a description   of the bound state similar to pNRQCD,
making apparent that the zero  order problem is the Schr\"odinger equation,
 we can still count on  another scale separation. 
 Let us consider   bound states of two nonrelativistic particles and some light d.o.f., e.g. molecules in QED  or quarkonium 
  hybrids ($Q\bar{Q} g$ states) or tetraquarks  ($Q\bar{Q}   q\bar{q}$  states) in QCD:
electron/gluon fields/light quarks  change adiabatically in the presence of heavy quarks/nuclei.  The heavy quarks/nuclei interaction may be described at leading 
order in the nonrelativistic expansion by an effective static energy (or  potential) $E_\kappa$ between the static sources where
$\kappa$ labels different excitations 
of the light degrees of freedom. 
 A plethora of states  can be  built on each on the static energies $E_\kappa$  by solving the
corresponding Schr\"odinger equation.  
This picture corresponds to the Born-Oppenheimer (BO) approximation \cite{Juge:2002br,Braaten:2014qka,Braaten:2014ita}.
Starting from pNRQED/pNRQCD the BO approximation can be made 
rigorous and cast into a suitable EFT called Born-Oppenheimer EFT (BOEFT)
\cite{Berwein:2015vca,Brambilla:2017uyf,Oncala:2017hop,Soto:2020xpm,Brambilla:2019jfi,Brambilla:2018pyn,Brambilla:2019esw}
which exploits the hierarchy of scales
$\lQ \gg mv^2$, $v$ being the velocity of the heavy quark.

\begin{figure}[!tbp]
 \begin{minipage}[b]{0.8\textwidth}
    \includegraphics[width=\textwidth]{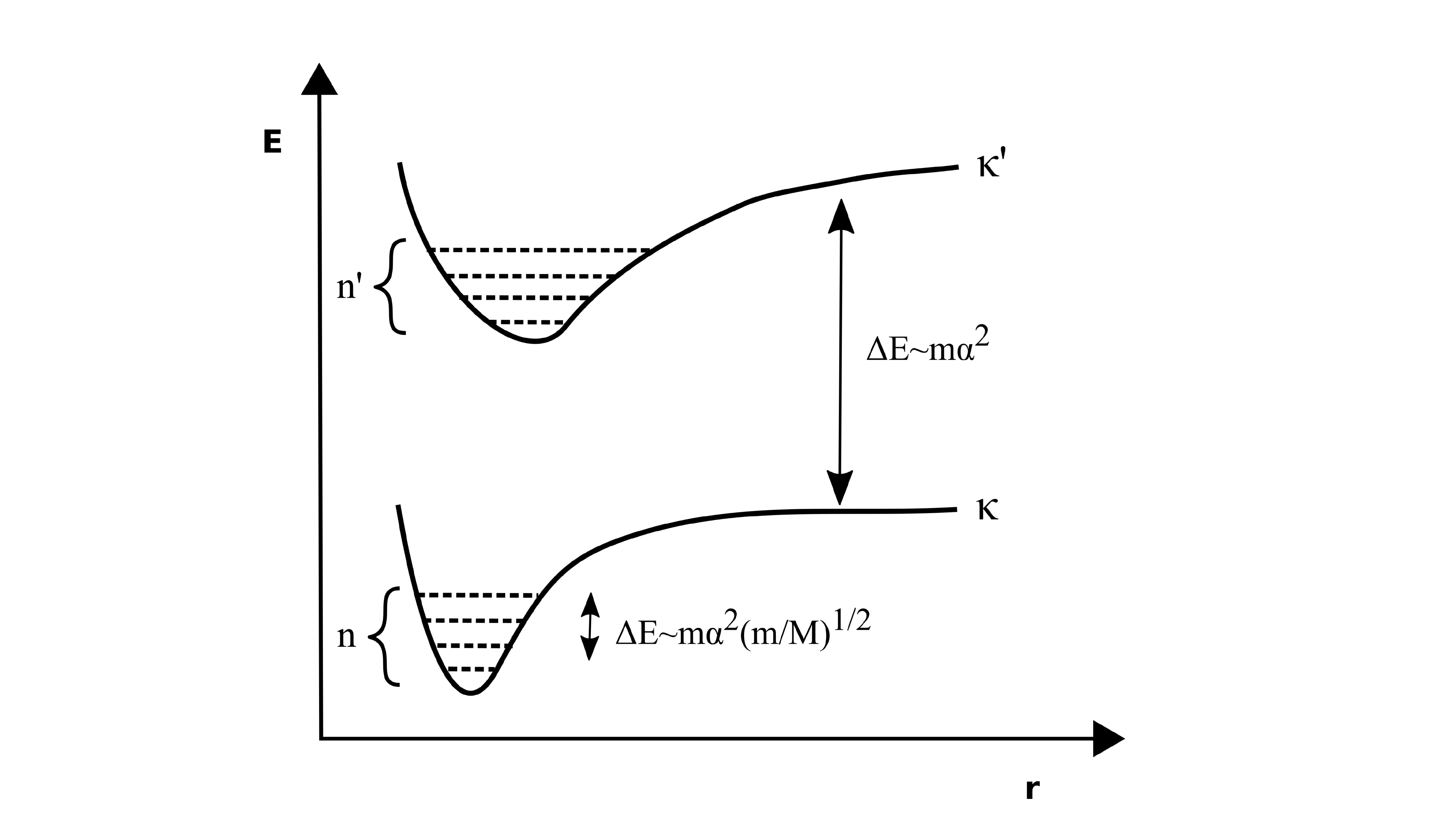}
    \vskip 0.2truecm  
    \caption{Pictorial view of electronic static energies in QED, labelled by a collective
    quantum number $\kappa$.} \label{bo1}
     \end{minipage}
\hfill
  \begin{minipage}[b]{0.8\textwidth}
    \includegraphics[width=\textwidth]{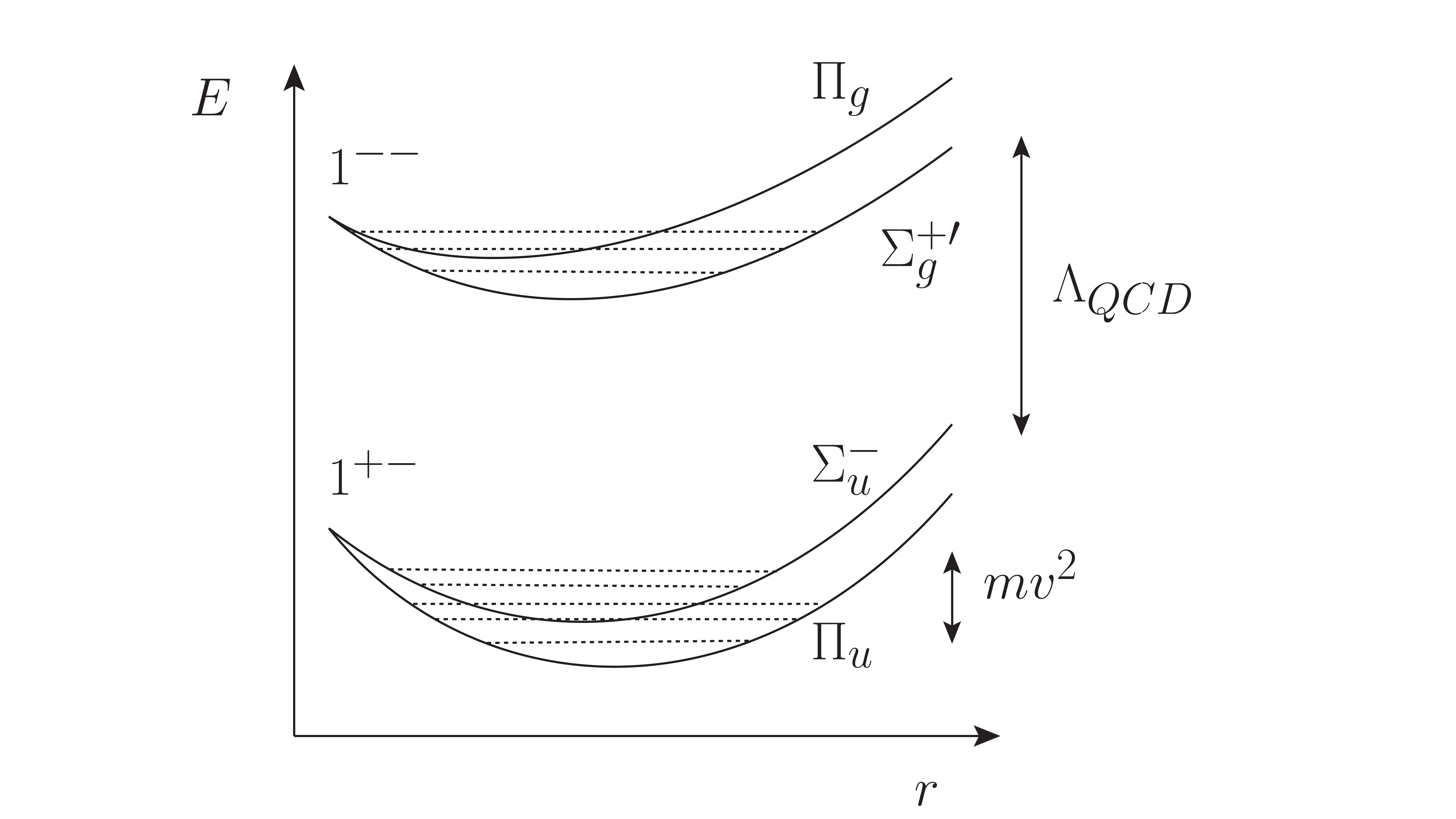} 
    \caption{Pictorial view of the gluonic (or hybrid) static energies, $E_\Gamma$, in QCD. The collective quantum
      number $\kappa$ has been detailed  in $\Lambda^\sigma_\eta$ as explained in the section  on strongly coupled pNRQCD. It corresponds to the actual
    lattice results in Fig. 8.}
      \label{bo2}
 \end{minipage}      
\end{figure}

\begin{figure}[ht]
 \centering
  \begin{minipage}[b]{1.2\textwidth}
    \includegraphics[width=\textwidth]{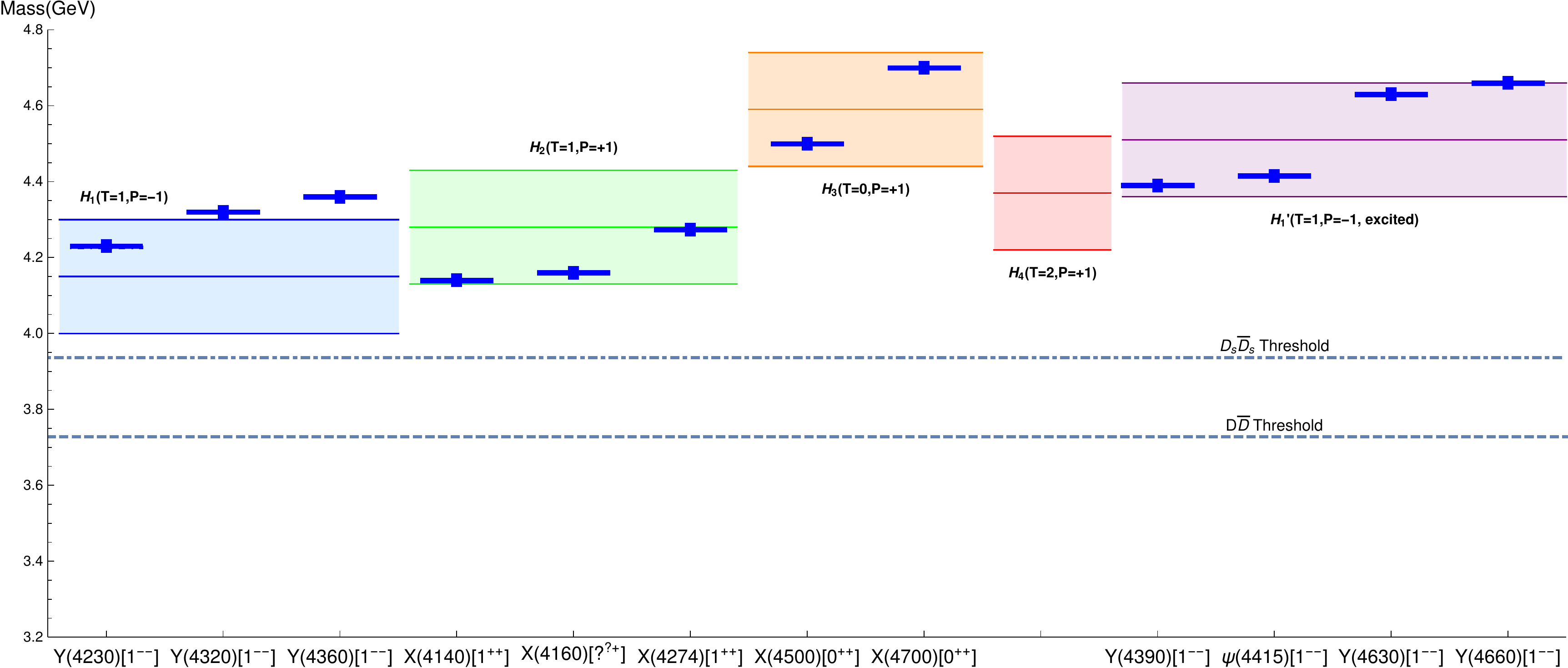}  
    \caption{Mass spectrum of neutral exotic charmonium states obtained by solving the BOEFT coupled Schr\"odinger equations.
      The neutral experimental states  that have matching quantum numbers are plotted in solid blue lines.
      In the figure $T$ stay for the total angular momentum. $H_1^\prime$ is the first $H_1$ radial excitation of $H_1$.
      The multiplets have been plotted with error bands corresponding to a gluelump mass uncertainty of 0.15 GeV.
      Figure taken from \cite{Brambilla:2019esw}.} \label{figexp}
     \end{minipage}
\end{figure}

In  \cite{Berwein:2015vca} we have obtained the BOEFT that describes hybrids. In particular we have obtained the static potentials 
and the  set of coupled   Schr\"odinger equations, solved them and produced all the hybrids multiplets, see Fig. \ref{figexp}, for the case
of the two first static energies $\Sigma_u^-$ and $\Pi_u$. Such static energies   are degenerated  at short distance where the cylindrical
symmetry gets subdue to a $O(3)$ symmetry and are then labelled by the quantum number of a  gluonic operator  $1^{+-}$ called a gluelump.
The hybrid static energies are described by a repulsive octet potential plus the gluelump mass in the short distance limit  and the $O(3)$ symmetry is broken 
at order $r^2$ of the multipole expansion.  In the long distance regime the static energies display a linear r behaviour.
The gluelump correlator can be calculated on the lattice to determine the  gluelump mass.
It is depending on the scheme used (the scheme dependence cancels against the analogous dependence in the quark  mass and in the octet static potential) but it is of the order of 800 MeV.
The hybrid multiplets $H_i$ are constructed from the solution of the Schr\"odinger equations  in correspondence of their $J^{PC}$
quantum numbers. The coupling between the different Schr\"odinger equations is induced by a nonadiabatic coupling, known in the Born Oppenheimer description of diatomic molecules, induced by
the noncommutation between the kinetic term and a projector of the cylindrical symmetry in the BOEFT lagrangian.

The degeneracy of the static energies at small distance  induces a  phenomenon called $\Lambda$ doubling,  removing the degenerations between
multiplets of opposite parity. This phenomenon is known in molecular physics but with smaller size. This and the structure of the multiplets differ 
from what is obtained in models for the hybrids, cf.  \cite{Berwein:2015vca}.  We used lattice input on the hybrid static energies and on the
gluelump mass.
 
  In \cite{Brambilla:2019jfi,Brambilla:2018pyn}  we obtained the spin dependent potentials at order $1/m$ and $1/m^2$ in the quark mass expansion and we could calculate all
  the hybrids spin multiplets. Notice that on one hand it is seldom to find the spin interaction considered in models, on the other hand it would be different.
  In fact  the $1/m$ contributions couple the angular momentum of the gluonic excitation with the total spin of the heavy-quark-antiquark pair.
These operators are characteristic of the hybrid states and are absent for standard quarkonia. 
Among the $1/m^2$ operators  besides the standard spin-orbit, total spin squared, and tensor spin operators respectively, which appear for standard quarkonia,
three novel operators appear. So interestingly, differently from the quarkonium case, the hybrid potential gets a first contribution already at order $\Lambda^2_{\rm QCD}/m$.
Hence, spin splittings are remarkably less suppressed in heavy quarkonium hybrids than in heavy quarkonia: this will have a
notable impact on the phenomenology of exotics.
We extracted the nonperturbative low energy correlators appearing in the factorization fixing them on lattice data on the masses 
of charmonium hybrids and we could then predict all the bottomonium hybrids spin multiplets, that are more difficult to evaluate on the lattice.
In this same framework it is also possible to calculate hybrids decays and quarkonium/hybrids mixing  \cite{Oncala:2017hop,Abi2021}.
 The BOEFT may be used to describe also tetraquarks, double heavy baryons and pentaquarks \cite{Brambilla:2017uyf,Soto:2020xpm}.
 In the case of tetraquarks,  a necessary input of the theory is the calculation of the 
lattice  generalized Wilson loops with appropriate symmetry and light quark operators, so that besides the quantum number
$\kappa$ also the isospin quantum numbers  $I=0, 1$ have to be considered.
The BOEFT approach reconciles  the different pictures of exotics based on tetraquarks, molecules, hadroquarkonium \dots
In fact in the
plot of a static energy as a function of $r$ for a state with $Q\bar{Q} q \bar{q} $ or $Q\bar{Q}  g $ we will have different regions: for 
short distance a hadroquarkonium picture would emerge, then a tetraquark (or hybrid) one and when passing the the heavy-light   mesons 
line,  avoided cross level effects  should have to be taken into account and a molecular  picture would emerge.
However QCD would dictate, through the lattice correlators and the BOEFT characteristics and power counting, which structure would dominate 
and in which precise way.  In addition production and suppression in medium may be described in the same approach \cite{Brambilla:2021abf, Brambilla:2020qwo}.
Production of hybrid states have been studied in  \cite{Chiladze:1998ti,Petrov:2005tp}.

\section{\textit{pNRQCD at finite T, open quantum system and quarkonium in medium}}

Quarkonium is a special probe also for deconfinement, besides confinement.
A prediction of QCD is that  at a certain value of temperature (or energy density) 
hadronic matter undergoes a transition to a deconfined state of quarks and gluons
called the quark-gluon plasma (QGP). Lattice QCD have shown that 
such transition takes place at a critical temperature  around 150 MeV \cite{Karsch:2001vs,Borsanyi:2020fev}.
Experimentally heavy ion collisions make it possible to study strongly interacting matter
under extreme conditions in the laboratory, recreating the QGP that 
should have primordially existed microseconds after the Big Bang.
These experiments have therefore a great importance  also for cosmology 
and allows to investigate the nuclear matter phase diagram, i.e. how nuclear matter 
change by varying the temperature and the chemical potential, see \cite{Brambilla:2014jmp, Andronic:2015wma}.
In heavy ions experiments at the LHC at CERN and at the RHIC at BNL 
the produced matter is characterized by small net baryon densities and high temperature.
 \cite{Karsch:2001vs}.
Heavy quarks are good probes of the QGP.
They are produced   at the beginning of the collision  and remain up to the end.
The heavy quark mass  $m$ introduces a large scale,
whose contribution may be factorized and computed in
perturbation theory. Low-energy scales  are sensitive  to the temperature   $T$  and even if
nonperturbative they  may be accessible via lattice calculations (for what concerns equilibrium physics).
As we discussed, quarkonia are special  hard probes as they are multi-scale systems.
In  the hot QCD medium also the thermal  scales of the Quark Gluon Plasma  (QGP) are emerging:
the scale related to the temperature $(\pi) T$, the Debye mass $m_D \sim g T$ related to the (chromo) electric screening and the scale  $g^2 T$
related to the  (chromo)magnetic screening. In a weakly coupled plasma the scales are separated and
hierarchically ordered, in a strongly coupled plasma $ m_D \sim T$.
To address QCD at finite T calculations, EFTs have been developed too to resum contributions related to 
the thermal scales and to address IR sensitivities. In real time the Hard Thermal Loop EFT (HTL)  \cite{Braaten:1989mz,Braaten:1991gm}
is taking care of integrating  out the temperature scale.

Heavy quarkonium dissociation  has been proposed long time ago as a clear probe of the quark-gluon plasma 
formation in colliders through the measurement of the dilepton decay-rate signal- In  \cite{Matsui:1986dk}
this was related to the screening of the quark-antiquark interaction due to 
Debye mass and it was suggested that this would have manifested in an exponential screening term 
$\exp (-m_D r)$ in the static potential. One of the key quantities  measured in experiments is the nuclear modification factor 
 $R_{AA} =Y(Pb Pb)/ (N_{coll} Y(pp)) $   where $Y(Pb Pb)$ and $Y(pp)$ are the quarkonium yield in PbPb and in pp collisions respectively.
 $R_{AA}$ is a measure for the difference in particle production in pp and nucleus-nucleus collisions.
Since higher excited quarkonium states have larger radius then  the expectation was
that, as the temperature increases, quarkonium would dissociate subsequently from the higher to the lower 
states giving origin to sequential melting.
In order to study quarkonium properties in a thermal bath at a temperature $T$, 
the quantity to be determined is the quarkonium potential $V$   which 
dictates, through  the Schr\"odinger equation
the real-time evolution of the wave function of a $Q\bar Q$ pair in the medium.
This  has been investigated for years
using many phenomenological assumptions,  spanning from the internal energy to the  free energy,  either the average free energy or the singlet one
which is gauge dependent.
pNRQCD has given us  the possibility to define in  QCD what is this potential:
 it is the matching coefficient   of the EFT that results from the integration of all the scales above the scale of the binding energy.
In a series of papers \cite{Brambilla:2008cx,Brambilla:2010vq,Brambilla:2013dpa,Brambilla:2011sg,Escobedo:2008sy,Brambilla:2020esl}
a pNRQCD at finite $T$ description has been constructed. 
One has to proceed integrating out all scales up to the binding energy and if the temperature is higher than the binding energy then 
also the temperature has to be integrated out using HTL EFT.
 When $T$ is bigger than the energy, the potential depends on the temperature, otherwise not.
 Thermal effects appear in any case in the nonpotential contributions to the energy levels.
 We assumed that the bound state exist for $T\ll m$ and $1/r \ge m_D$, we worked in the weak coupling limit and we consider 
all possible scales hierarchies  \cite{Brambilla:2008cx}.
We found that the thermal part of the potential has a real part (roughly described by the free energy)
and an imaginary part. The imaginary part comes from two effects:
the Landau damping \cite{Laine:2006ns,Escobedo:2008sy,Brambilla:2008cx}, an effect
existing also in QED, and the singlet to octet transition, existing
only in QCD \cite{Brambilla:2008cx}. Which one dominates depends on the ratio  between $m_D$ and $E$.
In the EFT one could show that the imaginary part of the potential related to the Landau damping comes
from inelastic parton scattering \cite{Brambilla:2013dpa} and the singlet to octet transitition from gluon dissociation \cite{Brambilla:2011sg}.
The existence of the imaginary part, first realized in \cite{Laine:2006ns}
 changed our paradigm for quarkonium suppression
as the state  dissociates  for this reason well before that the conventional screening becomes active
\cite{Laine:2006ns,Escobedo:2008sy}.
The pattern of thermal corrections is pretty interesting  \cite{Brambilla:2008cx}: when $T < E$ thermal corrections are
only in the energy; for $T> 1/r, 1/r>m_D$ or $1/r >T>E$ there is no exponential screening and $T$ dependent power like
corrections appear; if  $T> 1/r, 1/r\sim m_D$ we have exponential screening but the imaginary part of the static potential is
already bigger than the real one and dissociation already happened.
Once the potential has been calculated, the EFT gives the systematic framework to obtain the thermal energies: in \cite{Brambilla:2010vq}
it was performed the first QCD calculation of the thermal contributions to the  $\Upsilon(1S)$  mass and width
at order $m \alpha_s^5$ at LHC below the dissociation temperature
of about 500 MeV. This calculation is very important because it gives the parametric $T$ dependence of this observables. The width goes linear
in $T$ in the dominant term and this has been confirmed by lattice calculations of the spectrum \cite{Aarts:2011sm}.
These findings in the EFT in perturbation theory have inspired many subsequent nonperturbative calculations of the static potential
at finite $T$. In particular the Wilson loop at finite $T$ has been calculated on the lattice
\cite{Rothkopf:2011db,Rothkopf:2019ipj}
finding hints of a large imaginary parts. These calculations are pretty challenging and refining of the extraction methods are currently in elaboration
\cite{Bala:2021fkm}. 

Free energies defined by Polyakov loop correlators have been always very prominent in QCD at finite $T,$ see e.g. the reviews
\cite{Bazavov:2020teh,Ghiglieri:2020dpq}. 
 The Polyakov loop correlators of a single heavy quark and of a quark-antiquark pair  have been calculated  both in perturbation theory using pNRQCD to resum scales contributions
in   \cite{Berwein:2015ayt,Berwein:2013xza,Berwein:2017thy,Brambilla:2010xn}
and on the lattice to obtain these quantities fully nonperturbatively in \cite{Bazavov:2016uvm,Bazavov:2018wmo}.
In particular:  the Polyakov loop has been computed up to order $g^6$,
the (subtracted) $Q\bar{Q}$ free energy has been computed at short distances up to
corrections of order $g^7(rT)^4$, $g^8$, the (subtracted)  $Q\bar{Q}$
free energy has been computed at screening distances up to
corrections of order $g^8$; the singlet free energy has been computed at short distances up to corrections of
order $g^4(rT)^5$, $g^6$; the singlet free energy has been computed at screening distances up to
corrections of order $g^5$ \cite{Berwein:2015ayt,Berwein:2017thy,Brambilla:2010xn}.
From the lattice simulations and from comparison to the perturbative results some 
interesting information can be obtained \cite{Bazavov:2016uvm,Bazavov:2018wmo}: lattice calculations
are consistent with weak-coupling expectations in the regime of application
of the weakly coupled resummed perturbation theory which confirms the predictive power of the EFT;
the crossover temperature to the quark-gluon plasma  is $153+ 6.5- 5$ MeV  as 
extracted from the entropy of the Polyakov loop;  the screening sets in at $rT \sim 0.3-0.4$ (observable dependent),
consistent with a screening length of about $1/m_D$; asymptotic screening masses are
about $2m_D$ (observable dependent); the first determination of the color octet $Q\bar{Q}$
free energy has been obtained \cite{Bazavov:2018wmo}.
The free energies turn out not to be the objects to be used as a potential
in the Schr\"odinger equation even if the singlet  free energy may provide a good approximation of the real part of the static  potential.

These are all results in thermal equilibrium. However,  the evolution of quarkonium in the QGP is an
out of equilibrium process in which many effects enter: the hydrodynamical evolution of the plasma and the  production, dissociation 
and regeneration of quarkonium in the medium, to quote the most prominent ones.
It is necessary therefore to introduce an appropriate
framework to describe the real time nonequilibrium evolution of quarkonium in the QGP medium.
This has been realized in  \cite{Brambilla:2016wgg,Brambilla:2017zei,Brambilla:2019tpt}  where an open quantum
system (OQS) framework rooted in pNRQCD at finite $T$  has been developed that is fully quantum, conserves the
number of heavy quarks and considers both color singlet and color octet quarkonium degrees of freedom.
For a review of open quantum system approach see \cite{Akamatsu:2020ypb,Akamatsu:2014qsa,Sharma:2019xum,Yao:2021lus} and references therein.

We consider the density matrix associated to our system.
We distinguish between the environment (QGP), characterized by the scale $T$,  and the subsystem of system (quarkonium) characterized
by the scale $E$. We identify the inverse of E with the intrinsic time scale of the subsystem: $\tau_S\sim 1/E$
and the inverse of $\pi T$ with the correlation time of the
environment: $\tau_E\sim 1/(\pi T)$. If the medium is in thermal equilibrium,
or locally in thermal equilibrium, we may understand $T$ as a temperature, otherwise is
just a parameter. The medium can be strongly coupled.
Then we trace the density matrix over the environment and we are left with a color singlet and 
color octet density matrix that can be written in terns of the pNRQCD fields, working in the close time path formalism.
The evolution of the system is characterized by a relaxation time $\tau_R$   that  is estimated by the inverse
of the color singlet self-energy diagram in pNRQCD at finite $T$.
We select quarkonia states with a  small radius (Coulombic) for which $1/r \gg \pi T, \Lambda_{QCD}$ and we
consider $\pi T \gg E$.
In this framework ,in~\cite{Brambilla:2017zei}, a  set of  master equations  governing the
time  evolution of  heavy quarkonium  in a  medium can be   derived.
The equations  follow  from  assuming  the  inverse  Bohr  radius  of  the
quarkonium  to  be  greater  than  the energy  scale  of  the  medium,  and model  the  quarkonium as  evolving in  the
vacuum up  to a time  $t=t_{0}$ at  which point interactions  with the medium begin.
The equations express the time evolution of the density
matrices  of the  heavy quark-antiquark  color singlet,  $\rho_s$, and
octet  states, $\rho_o$,  in  terms  of the  color  singlet and  octet
Hamiltonians, $h_s  = {\bf p}^2/M  - C_F\alpha_s/r  + ...$ and  $h_o =
{\bf p}^2/M + \alpha_s/(2N_cr) +  ...$, and interaction terms with the
medium, which, at order $r^2$  in the multipole expansion, are encoded
in  the self-energy  diagrams of the EFTs.
 These interactions account for  the mass shift of  the heavy $Q\bar{Q}$
pair induced by the medium, its decay width induced by the medium, the
generation    of   $Q\bar{Q}$ color   singlet    states   from
   $Q\bar{Q}$  color octet states interacting with the medium and the
generation of   $Q\bar{Q}$  color  octet states from $Q\bar{Q}$
(color singlet or octet) states interacting with the medium (recombination terms).
Both Landau damping and singlet to octet effects are contained in the self energy.
The leading order  interaction between  a heavy  $Q\bar{Q}$
  field and the  medium   is  encoded  in   pNRQCD  in  a   chromoelectric  dipole
interaction, which  appears at  order $r/m^0$  in the  EFT  Lagrangian.
The approach gives us  master equations, in general non Markovian,
for the out of equilibrium evolution of the color singlet and color octet matrix densities.
The system is in non-equilibrium because through interaction with the environment
(quark gluon plasma) singlet and octet quark-antiquark states continuously transform in
each other although the total number of heavy quarks is conserved.

\begin{figure*}[ht]
	\begin{center}
		\includegraphics[width=0.43\linewidth]{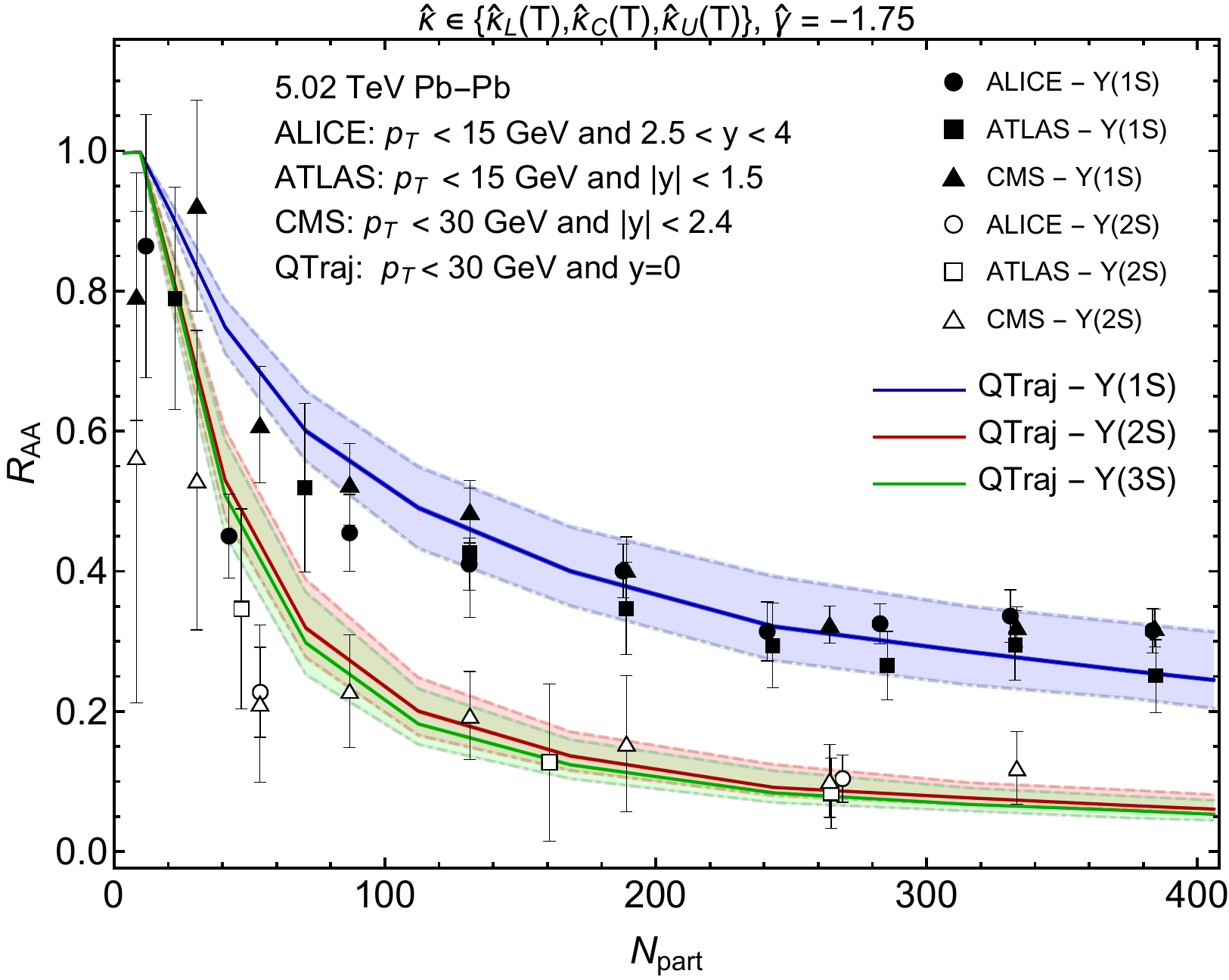}  \hspace{1cm}
		\includegraphics[width=0.43\linewidth]{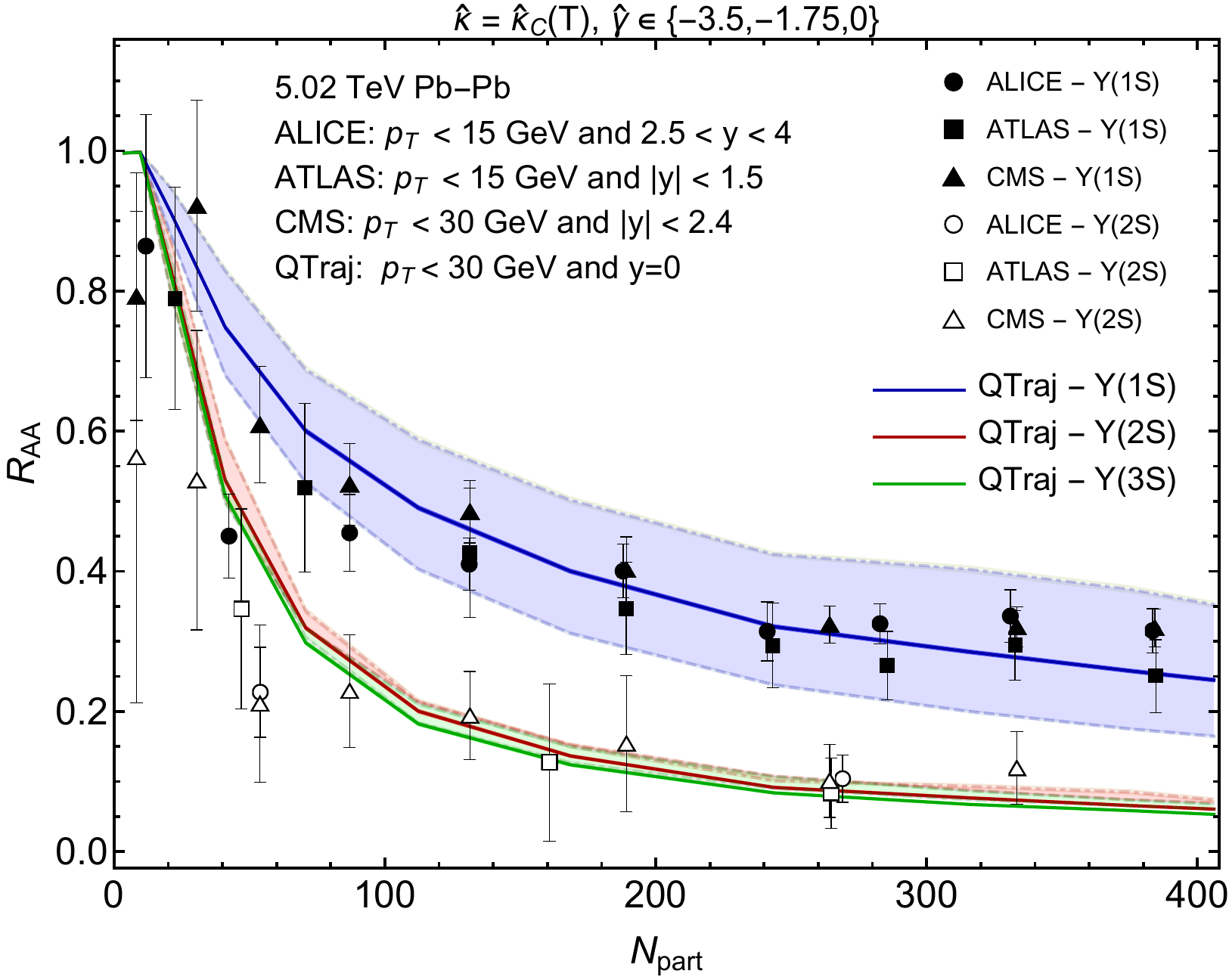}
	\end{center}
	\caption{The nuclear modification factor $R_{AA}$ of the $\Upsilon(1S)$, $\Upsilon(2S)$, and $\Upsilon(3S)$ as a function of $N_{{part}}$ compared to experimental measurements from the ALICE~\cite{Acharya:2020kls}, ATLAS~\cite{ATLAS5TeV}, and CMS~\cite{Sirunyan:2018nsz} collaborations.
		The bands in the theoretical curves indicate variation with respect to $\hat{\kappa}(T)$ (left) and $\hat{\gamma}$ (right).
		The central curves represent the central values of $\hat{\kappa}(T)$ and $\hat{\gamma}$, and the dashed and dot-dashed lines represent the lower and upper values, respectively, of $\hat{\kappa}(T) \equiv
		\kappa/T^3$ and $\hat{\gamma}\equiv \gamma/T^3$ ,  see text for the definition of these parameters. Figure taken from \cite{Brambilla:2021wkt} . 
	}
	\label{fig:raa_vs_npart}
\end{figure*}

Assuming that any energy scale in the medium is
larger than the heavy   $Q\bar{Q}$
 binding energy $E$, in particular that
$\tau_R \gg \tau_E$, we obtain a Markovian evolution while  the chosen hierarchy of scales
implies $\tau_s\gg \tau_E$ qualifying the regime as quantum Brownian motion.
In this situation we can reduce the general master equation to a Linblad form.
In this case the properties of the QGP are encoded in two
transport coefficients: the heavy quark momentum diffusion coefficient, $\kappa$, and its  dispersive counterpart
$\gamma$ which are given by time integrals of appropriate gauge invariant correlators at finite $T$ given 
by the integral of gauge invariant finite T  QCD correlators of chromoelectric fields:
\begin{eqnarray}
	\kappa&=&\frac{g^{2}}{6N_{c}}\int^{\infty}_{0}\mathrm{d}s~\Big \langle \left\{ E^{a,i}(s,{\bf 0}),E^{a,i}(0,{\bf 0}) \right\} \Big \rangle, \label{eq:kappa_def} \\
	\gamma&=-i&\frac{g^{2}}{6N_{c}}\int^{\infty}_{0}\mathrm{d}s~\Big \langle \left[ E^{a,i}(s,{\bf 0}),E^{a,i}(0,{\bf 0}) \right] \Big \rangle. \label{eq:gamma_def}
\end{eqnarray}
They come from the pNRQCD self energies that in this regime could be factorized between this contribution and the bound state 
dependent part. In the case of a nonperturbative QGP, these objects are nonperturbative and should be evaluated on the lattice at a given temperature in an extended window of temperatures
\cite{Brambilla:2020siz}. In this way one could use lattice to give input to a nonequilibrium calcultation.
In a series of papers \cite{Brambilla:2021wkt,Brambilla:2020qwo}
 we have solved the Linblad  equation  using the highly efficient quantum trajectory method
  and realistically implementing   a 3+1D
dissipative  hydrodynamics code \cite{Omar:2021kra}, and we have obtained  the  bottomonium  ($\Upsilon(1S), \Upsilon(2S), \Upsilon(3S)$) nuclear
modification factors  and the anisotropic flows,
see Fig. \ref{fig:raa_vs_npart} for a comparison between  LHC data and
 our  results on $R_{AA}$. 
  The computation does not  rely on any free parameter, as it  depends only on the two
transport  coefficients  that   have  been  evaluated independently  in
lattice QCD. Our  final results, which include late-time  feed down of
excited  states, agree  well with the  available data from LHC  5.02 TeV
PbPb collisions.  Notice that the bands in the Fig. \ref{fig:raa_vs_npart}
depend on the undetermination with which the nonperturbative transport coefficients $\kappa$ 
and $\gamma$, that are properties of the QGP,  are presently known. More precise experimental data can help to narrow down 
their range.
In this way we can use quarkonia with small radii
as a diagnostic and investigation tool of the characteristics of the strongly coupled QGP.
To apply the same description to charmonium would require to modify the master equations considering 
terms beyond leading order in the quarkonium density. 
Further work in a similar approach at the level of the Boltzmann equation has been done in 
\cite{Yao:2020kqy,Yao:2018nmy}.

\section{\textit{Outlook}}

The great progress of the last few decades in the construction of nonrelativistic
effective quantum field theories and the progress in lattice QCD calculation allows
to treat  heavy quark bound states systematically in QCD. In this way quarkonium
becomes a golden probe of strong interactions at zero and finite temperature.

Moreover, quarkonium is the prototype of a nonrelativistic multiscale system.
Systems of such kind are  ubiquitous in matter and in  fields of physics,
from  particle to  nuclear  physics, to condensed  matter and  to
astro and cosmological  applications and play a key role in
several open challenges at the frontier of particle physics.
Therefore, all the progress obtained in this framework holds the promise
to have an impact in a number of relevant contemporary problems.

For example. combining the techniques of nonrelativistic effective field theories
and the open quantum framework, we can address the nonequilibrium evolution of heavy
dark matter pairs in the early  universe.  On the other hand the XYZ exotics bear
similarities to atomic and molecular physics, and results obtained in BOEFT can be used in those
fields \cite{Brambilla:2017uyf,Brambilla:2017ffe} as well as maybe extended to the consideration of strongly
correlated system in condensed matter at finite temperature.

In summary, the physics of heavy quarks not only has been and is extremely important for our advancement in nuclear and particle
physics, but has the promise to impact also many other nearby fields.

%
%
%
%

\end{document}